\documentclass[preprint,preprintnumbers,amsmath,amssymb,nofootinbib]{revtex4}
\usepackage{graphicx,color}
\newcommand\nn{\nonumber \\}

\def\sla{\slash{\!\!\!}}

\newcommand{\lk}{\left(}
\newcommand{\rk}{\right)}
\newcommand{\ltk}{\left\{}
\newcommand{\rtk}{\right\}}
\newcommand{\ldk}{\left[}

\newcommand{\rdk}{\right]}
\newcommand\beq{ \begin{eqnarray} }
\newcommand\eeq{ \end{eqnarray} }

\begin{document}

%\preprint{}
\title{BEC-polaron gas in a boson-fermion mixture : 
a many-body extension of Lee-Low-Pines theory}
\author{Eiji Nakano}
\email{e.nakano@kochi-u.ac.jp}
\affiliation{Department of Physics, Kochi University, 
Kochi 780-8520, Japan}
\author{Hiroyuki Yabu} 
\email{yabu@se.ritsumei.ac.jp}
\affiliation{Department of Physics,
Ritsumeikan University, Kusatsu 525-8577, Siga, Japan}

\date{\today}

\begin{abstract}
We investigate the ground state properties of the gaseous
mixture of a single species of bosons and fermions at
zero temperature,
where bosons are major in population over fermions, 
and form the Bose-Einstein condensate (BEC). 
The boson-boson and boson-fermion interactions are assumed to be weakly
repulsive and attractive respectively, 
while the fermion-fermion interaction is absent due to the
Pauli exclusion for the low energy $s$-wave scattering.
We treat fermions as a gas of polarons dressed with
Bogoliubov phonons, 
which is an elementary excitation of the BEC,
and evaluate the ground state properties  
with the method developed by Lemmens, Devreese,
and Brosens  (LDB) originally for the electron polaron gas, 
and also with a general extension of the 
Lee-Low-Pines theory for many-body systems (eLLP), 
which incorporates the phonon drag effects as in the original LLP theory. 
The formulation of eLLP is developed and discussed in the present paper. 
The binding (interaction) energy of the polaron gas is calculated 
in these methods, and shown to be finite (negative) 
for the dilute gas of heavy fermions with attractive 
boson-fermion interactions,
though the suppression by the many-body effects exists.
\end{abstract}

%\pacs{}
%\keywords{}

\maketitle

%\tableofcontents

\section{Introduction}
Elementary excitations in condensed matter physics are
important degrees of freedom to understand
various phenomena from the many-body point of view \cite{Mahan1}.
Polarons are such excitations observed for electrons
conducting in polar crystal environments,
where electrons are dressed with excited phonons, 
and drag due to interaction with them \cite{Mahan1,polaronreview1}.
Theoretical development for the electron polarons
is originated in Landau and Pekar's works \cite{Landau1}, and 
later the modern concept has been established by 
Fr\"ohlich \cite{FPZ1} for the electron-phonon systems; 
such systems have been studied by various methods 
including the mean-field type approach by Lee, Low, and Pines (LLP) 
\cite{LLP1}, 
Feynman path-integral approach \cite{Feynman1,OpticalAbsoption1},
many-body Green function methods \cite{Green1,Green2,Devreese1,Devreese2},  
and models for small polarons have also been developed 
\cite{Yamashita1,Yamashita2,Yamashita3}. 

Recently, the BEC-polarons attract interests, 
which occurs in the Bose-Einstein condensate (BEC) of optically trapped ultra-cold atoms 
\cite{Pitaevskii1,PethickSmith1}: 
a single atomic impurity immersed in the BEC 
dressed with BEC Bogoliubov phonons 
\cite{Cucchietti1,Sacha1,Tempere1,Casteels1,
Shchadilova1,Christensen1}. 
Also, there are studies on atomic polarons 
in the environment of interacting Fermi gas 
\cite{Chevy1,Schirotzek1,Kohstall1,Koschorreck1,Vlietinck2,Massignan1,Yi1}. 
Since they are conceptually similar to the electron
polaron with crystal phonons,  
conventional methods mentioned above
are applied to study them theoretically. 

The experimental advantages in treating these atomic impurities 
in the BEC is that, 
because of the controllability of the systems 
using the change of optical trap of the system 
and the interatomic interactions using the Feshbach resonances,
the various properties of the BEC-polaron systems 
can be observed in various situations: 
the mobility, the damping rate, 
and the binding energy of impurities.
For instance, the direct observations of the energy of the BEC-polaron 
to the bare impurity is proposed from the radio-frequency absorption between two hyperfine states,
which are dressed (polaron-like) and undressed (bare-impurity-like) 
with Bogoliubov phonons \cite{Shashi1}. 
An another experimental possibility is 
to trace the position of the BEC-polaron 
in the optical traps \cite{Hohmann1}, and 
to tune the interaction intensity between impurities and the condensate 
by external lasers \cite{Compagno1}.
\footnote{Recently attractive and repulsive BEC-polaron systems have been observed experimentaly \cite{Jrgensen1}.}

In contrast to the conventional electron-phonon system 
with the Coulomb interaction, 
the effective interaction between atomic impurity and 
Bogoliubov phonons in the BEC can be tuned from weak to 
strong couplings,
including the unitarity limit 
where two atoms start to form a molecular bound state 
\cite{Zwerger1,Storozhenko1,Fratini,Ludwig1,Guidini1}. 
For studies of the single polaron in such strongly-coupled systems,  
more advanced non-perturbative methods are needed 
\cite{Rath1,Vlietinck1,Grusdt1,Grusdt3,Yin1,Grusdt2,Levinsen2}.  
Especially, in the region around the unitarity limit, 
the non-perturbative renormalization group method has figured out 
the spectral properties of Fermi polarons 
in the polaron-molecule crossover regime \cite{Schmidt1}. 
Recently, a quantum Monte-Carlo method has been used 
for the microscopic description of impurities in the BEC, 
which includes regimes from weak to strong coupling constants \cite{Ardila2}.

The aim of the present paper is at studying the case 
where the number of fermionic impurities is increased to 
make a dilute Fermi gas in the BEC.  
For this purpose, 
we consider the gaseous mixture of single component bosons and fermions,
where immersed fermions are treated as a dilute gas of polarons 
interacting with the Bogoliubov phonons excited in the BEC; 
the interaction between fermion and boson (phonon) 
is assumed to be weakly attractive.
Thus the strong correlation effects 
such as boson-fermion pair fluctuations are irrelevant in the mixture. 

In this paper,  
we calculate the ground state properties of the system 
at zero temperature as a BEC-polaron gas, 
dressed with phonon clouds as in the single polaron treatment. 
To this end we first employ the unitary transformation method 
by Lemmens, Devreese, and Brosens (LDB) originally developed 
for the gas of electron-phonon polarons 
\cite{LDB1,Tempere4,Putteneers1},  
which has been applied to 
many-polaron systems in the BEC 
for bosonic and fermionic impurities 
in general situations with bare interactions among impurities \cite{LDB2,LDB3}. 
Then, we also develop the method using
the different unitary transformation,
which generalize the LLP theory of the single polaron to many-polaron systems, 
in order to incorporate the drag effect absent in LDB. 
In these methods, 
we evaluate the ground state energy of the many-polaron gas, 
and the single polaron properties in the gas, 
such as the binding energy per fermion and the effective mass, 
and calculate their dependence on the density and mass ratios of the fermion to the boson, 
and on the boson-fermion interaction strength. 
 
This paper is organized as follows:  
In section II, 
we introduce a low energy effective Hamiltonian for the boson-fermion mixture,
and implement the Bogoliubov approximation 
to obtain a Fr\"{o}hlich-type effective Hamiltonian. 
In section III, 
we study the single BEC-polaron system in the LLP theory, 
and show some properties of the solution.
In sections IV and V, 
we apply the LDB and the eLLP methods 
to the Fr\"{o}hlich-type effective Hamiltonian
obtained in section III,
and evaluate the ground state properties.  
We also compare the obtained results with those from the LLP theory    
for the single polaron in the appropriate limit. 
The last section is devoted to summary and outlook.  

%%%%%%%%%%%%%%%%%%%%%%%%%%%%%%%%%%%%%%%%%%%%
\section{Low energy effective Hamiltonian}
%%%%%%%%%%%%%%%%%%%%%%%%%%%%%%%%%%%%%%%%%%%%%
We consider the uniform system of the gaseous mixture 
consisting of single species of bosons and fermions.
In terms of the boson and fermion field operators,
$\phi(r)$ and $\psi(r)$,
the effective Hamiltonian of the system is
\beq
\mathcal{H}
&=& 
-\int_r \psi^\dagger(r)
\frac{\nabla^2}{2 m_f} \psi(r)
-
\int_r \phi^\dagger(r)
\frac{\nabla^2}{2 m_b} \phi(r)
\nn
&&
+g_{bf}\int_r \psi^\dagger(r)\psi(r) 
\phi^\dagger(r)\phi(r)
+\frac{1}{2}
g_{bb}\int_r
\phi^\dagger(r)\phi^\dagger(r)\phi(r)\phi(r)
%%%%%%%%%%%%%%%
\nn
&=& 
\sum_p \lk \xi_p a_p^\dagger a_p + \varepsilon_p
b_p^\dagger b_p \rk
+\frac{1}{V} \sum_{k,p,q} \left\{ g_{bf} a_{p+q}^\dagger
b_{k-q}^\dagger b_k a_p
+\frac{1}{2} 
g_{bb}  b_{p+q}^\dagger b^\dagger_{k-q} b_{k} b_{p} \right\},
\label{Hamil1}
\eeq
where the Fourier expansions,
$\phi(r)=V^{-1/2}\sum_p e^{ip\cdot r} b_p$ and 
$\psi(r)=V^{-1/2}\sum_p e^{ip\cdot r} a_p$, 
have been used 
with the discreet values of momentum $p$ in the volume $V$ 
(to be sent to infinity for the thermodynamic limit), 
and the free single-particle energies of bosons and fermions are  
$\varepsilon_p = \frac{p^2}{2m_b}$ and 
$\xi_p = \frac{p^2}{2m_f}$ respectively,
with the bare boson and fermion masses $m_b$ and $m_f$.
The creation and annihilation operators satisfies 
the commutation or anti-commutation relations: 
$\ldk b_k, b_p^\dagger \rdk = \delta_{k,p}$ and
$\left\{ a_k, a_p^\dagger \right\} = \delta_{k,p}$. 
Throughout this paper, 
we use the abbreviations $\int_r\equiv \int{\rm d}^3{r}$ and 
$\int_p \equiv \int{\rm d}^3{k}/(2\pi)^3$ 
for the real and momentum space integrals respectively, 
and use the natural unit $\hbar=c=1$. 

In the case of the mixture of the dilute gas, 
the boson-boson and boson-fermion coupling constants, 
$g_{bb}$ and $g_{bf}$, are represented 
by the s-wave scattering lengths, 
$a_{bb}$ and $a_{bf}$, respectively;
in the T-matrix approach, 
the relations are given by 
\cite{PethickSmith1}
\beq
\frac{m_{ij}}{2\pi a_{ij}}
=
\frac{1}{g_{ij}}+\int_p
\frac{1}{p^2/(2 m_i) +p^2/(2 m_j)},
\quad (i,j =b,f)
\label{Tmatrix}
\eeq
where $m_{ij}=\frac{m_im_j}{m_i+m_j}$ is the reduced mass of particles $i$ and $j$.
In the weak coupling regime, 
it becomes 
$g_{bb}=\frac{4\pi}{m_b}a_{bb}$ and
$g_{bf}=\frac{2\pi}{m_{bf}}a_{bf}$ for the boson-boson and boson-fermion interactions.
The above formulation is valid only for systems 
with a mean interparticle distance much larger 
than a typical size of particles $r_0$, 
which introduces the natural cutoff of $\sim 1/r_0$ in the momentum integral in Eq.~(\ref{Tmatrix}).
\subsection{Bogoliubov phonon of BEC}
In the mixture of the weak boson-boson repulsive and boson-fermion attractive interactions, 
we assume that all bosons are in the state of the Bose-Einstein condensation (BEC) at zero temperature, 
and the low-energy elementary excitation is primarily the Bogoliubov phonon.
Thus,  
keeping only terms including the condensation parts 
with zero momentum to the quadratic order of excitations in the boson sector, 
we obtain the Hamiltonian (Appendix~\ref{ApBog} for detail) :  
\beq
H_b
&=&
\sum_p \frac{p^2}{2 m_b}b_p^\dagger b_p 
+
\frac{1}{2V} g_{bb} 
\sum_{k,p,q}
b_{p+q}^\dagger b^\dagger_{k-q} b_{k} b_{p} 
%%%%%%%%%%%%%%%%%%%%%
\nn
&\simeq& 
\frac{1}{2} g_{bb} \frac{N_b^2}{V}
+\frac{1}{2}\sum_{q \neq 0} ( E_q -\varepsilon_q -g_{bb}n_0 )
+\sum_{q \neq 0} E_q C_q^\dagger C_q, 
\label{Bogol1}
\eeq
where $N_b$ is the boson total number, 
and $n_0 =N_0/V$ is the condensed-boson density obtained from the condensed-boson number $N_0$, 
which is approximated by $N_0 \sim N_b$ 
in the present system of the weak interactions and zero-temperature.   
The $C_q$ and $C_q^\dagger$ are the annihilation 
and creation operators of the Bogoliubov phonon 
with the excitation energy: 
\beq
     E_q=\sqrt{\varepsilon_q\lk \varepsilon_q + 2 g_{bb}n_0\rk},
\eeq
and they satisfy the commutation relations:  
$\ldk C_p,C_q^\dagger \rdk=\delta_{p,q}$, and others.   
The first and second $c$-number terms in the last line of (\ref{Bogol1}),
which correspond to the ground state energy of the pure Bosonic gas, 
are dropped in the remaining part of this paper.  

\subsection{Fr\"ohlich-type Hamiltonian of phonon-fermion system}
We also use the Bogoliubov approximation 
for the boson-fermion interaction (Appendix~\ref{ApBog}), 
and obtain the Fr\"ohlich-type Hamiltonian of the interacting fermion-phonon system 
from (\ref{Hamil1}) and (\ref{Bogol1}):
\beq
H&=&H_{f}+H_{b}+H_{int} \nn
 &=& \frac{1}{2 m_f}\int_r \nabla\psi^\dagger(r) \cdot \nabla\psi(r)
    +\sum_{q\neq 0} E_q C_q^\dagger C_q \nn
 & &+g_{bf} \frac{N_0 N_f}{V}+
     \int_r \psi^\dagger(r) \psi(r)  
     \sum_{q\neq 0}
     g_{q} \lk e^{-ir\cdot q} C_{q}^\dagger + e^{ir\cdot q} C_q\rk, 
\label{FrH}
\eeq
where we have replaced the fermion number operator $\int_r \psi(r)^\dagger \psi(r)$  
by the total fermion number $N_f$ of the mixture, 
and the Yukawa-type coupling constant $g_q$ for the fermion-phonon interaction is given by 
\beq
g_q =\frac{N_0^{\frac{1}{2}}}{V} g_{bf} (u_k-v_k)
    =\frac{N_0^{\frac{1}{2}}}{V} g_{bf} \sqrt{\frac{\varepsilon_q}{E_q}}. 
\label{YukawaC}
\eeq 
Noted that it includes the momentum dependent factor, 
$(u_k-v_k)=\sqrt{\frac{\varepsilon_k}{E_k}}$, 
stemming from the Bogoliubov transformation,
defined in (\ref{BogTransf}) in Appendix~\ref{ApBog}.

%%%%%%%%%%%%%%%%%%%%%%%%%%%%%%%%%%%%%%%%%%%%%%%%%%%%%%%%
\section{Single Polaron in BEC: Lee-Low-Pines theory}
%%%%%%%%%%%%%%%%%%%%%%%%%%%%%%%%%%%%%%%%%%%%%%%%%%%%%%%
In this section we review the single BEC-polaron system 
for weak/intermediate interaction regimes, 
and show some properties of the solution obtained in the LLP theory: 
the ground state energy, drag parameter $\eta$, and the effective mass 
which are already presented in e.g., \cite{Shashi1,Grusdt2}. 
In addition,  
we estimate the size of phonon cloud directly from the solution, 
and give validity conditions for the effective Fr\"{o}hlich Hamiltonian. 
These results in the LLP theory will be helpful as references  
in discussions on many-body BEC-polaron systems later. 
%
%%%%%%%%%%%%%%%
\subsection{LLP transformation}
%%%%%%%%%%%%%%
In the case of a single fermion impurity immersed in the BEC at the position $x$, 
the fermion density operator is represented as  
$\psi^\dagger(r)\psi(r)=\delta^{(3)}(r-x)$, 
and the Hamiltonian (\ref{FrH}) becomes
\beq
H &=& H_f+H_b+H_{int} \nn
  &=& -\frac{\nabla_x^2 }{2 m_f} 
     +\sum_{q \neq 0} E_q C_q^\dagger C_q 
     +\sum_{q \neq 0} g_q
      ( e^{-i x \cdot q} C_{q}^\dagger + e^{i x  \cdot q} C_q ) +g_{bf} n_0. 
\label{FrSingle}
\eeq
This Hamiltonian enables us to map the argument of the conventional electron-polaron 
onto the BEC-polaron, 
so in order to discuss the ground state properties of it 
we employ the Lee-Low-Pines (LLP) theory for relatively weak coupling regimes, 
in which some unitary transformations are utilized. 
These transformations are also used in extended forms for many-body systems of fermions 
in the later parts of the present paper. 

First, the unitary transformation $S(x)$ is defined by 
\beq 
S(x) = 
\exp\left[ -i x \cdot \sum_{q \neq 0} q C_q^\dagger C_q \right], 
\label{transS}
\eeq
which serves as a gauge transformation for the phonon operators:
\beq
S^{-1} C_q S =C_q e^{-i x \cdot q}, \quad
S^{-1} C_q^\dagger S =C_q^\dagger e^{i x \cdot q},
\eeq
and remove the phonon contribution from the total momentum operator:
\beq 
 S^{-1} \left\{ -i\nabla_x +\sum_{q \neq 0} q C_q^\dagger C_q \right\} S 
 =  -i\nabla_x. 
\eeq
It implies that the transformed momentum operator is that observed in the frame comoving 
with the impurity.
So the transformed Hamiltonian does not depend on the coordinate $x$:
\beq
\tilde{H}\equiv 
S^{-1}HS
&=&\frac{1}{2 m_f} 
\lk -i\nabla_x -\int_q q C_q^\dagger C_q\rk^2
+ \sum_{q \neq 0} E_q C_q^\dagger C_q
\nn
&&+ 
\sum_{q \neq 0}
g_q
\lk C_{q}^\dagger + C_q\rk
+g_{bf} n_0,    
\eeq
and it includes the phonon-phonon interaction term 
which does not exist originally in (\ref{FrSingle}). 
Since the transformed Hamiltonian $\tilde{H}$ 
commutes with the momentum operator of the impurity, 
we can replace the operator  $-i\nabla_x$ in (\ref{FrSingle})
with the $c$-number $P$ that is the momentum eigenvalue 
when we consider the plane-wave state $e^{i P \cdot x}$ for the impurity; 
consequently the parameter $P$ in the transformed Hamiltonian 
is the total momentum of the single polaron including that of the dressed phonon.
Now the problem reduces to solve the energy eigenvalue equation:
$\tilde{H}(P)|\Psi\rangle = E(P)|\Psi\rangle$.  

Second,  
as the ground state for the $\tilde{H}$ in the LLP theory, 
we take the state of the phonon cloud: 
$|\Psi \rangle =T |0 \rangle$ 
where $|0\rangle$ is the phonon vacuum state, 
and the unitary transformation operator $T$, which produces the phonon cloud, 
is defined by 
\beq
T =\exp\left[ \sum_{q\neq 0} \lk f_q C_q^\dagger-f_q^*C_q\rk \right].
\label{transT}
\eeq 
We should note that the state $| \Psi \rangle =T | 0 \rangle$ is a coherent state 
with the parameter $f_q$;
through the relation $f_q =\langle 0 | T^{-1} C_q T |0 \rangle$. 
The parameter $f_q$ is found to be the phonon momentum amplitude of the momentum $q$ in the state $|\Psi\rangle$,  
which is to be determined variationally from the minimum of the energy expectation value 
$\langle \Psi| \tilde{H}(P)|\Psi\rangle$.

Accordingly, in the LLP theory,
the ground state of a single polaron with momentum $P$ 
for the Hamiltonian $H$ is described by the product state: 
\beq
|x;P \rangle \equiv 
e^{i x \cdot P} ST| 0 \rangle
=
e^{i x \cdot P} U S|0\rangle
=
e^{i x \cdot P} U|0\rangle, 
\label{LLPgs}  
\eeq
where the unitary operator $U$ is defined as
$U=STS^{-1}= e^{Q(r)}$:
\beq
U=STS^{-1}= e^{Q(r)}
 =\exp\left[
\sum_{k \neq 0} 
(e^{-i k \cdot x } f_k C_k^\dagger
-f_k^* e^{i k \cdot x } C_k ) \right],  
\label{transU}
\eeq
and $S| 0 \rangle=| 0 \rangle$ has been used
in the derivation of (\ref{LLPgs}). 

\subsection{Ground state energy and drag parameter}
\begin{figure}[h]
  \begin{center}
    \begin{tabular}{cc}
\resizebox{70mm}{!}{\includegraphics{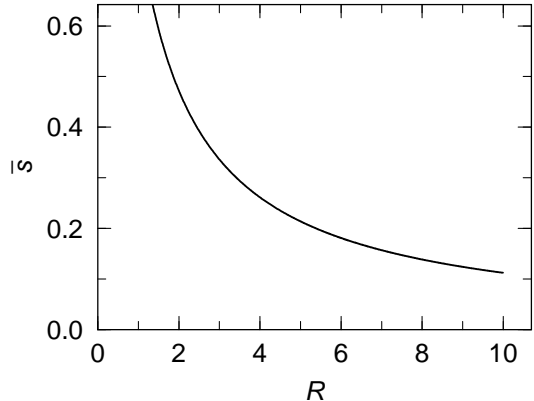}} &
\\
% \resizebox{50mm}{!}{\includegraphics{nemo2.eps}} \\
% \resizebox{50mm}{!}{\includegraphics{nemo2.eps}} &
% \resizebox{50mm}{!}{\includegraphics{nemo1.eps}} \\
    \end{tabular}
    \caption{Parameter $s$,  
    normalized by $s=\bar{s} a_{bf}^2n_0^{\frac{1}{2}}/a_{bb}^{\frac{1}{2}}$,  
    as a function of the mass ratio $R=m_f/m_b$. }
    \label{test1}
  \end{center}
\end{figure}

The energy expectation value with the ground state (\ref{LLPgs}) is calculated to be 
\beq
E_{pol}(P) &\equiv& \langle x;P | H |x;P \rangle 
\nn
&=&
\frac{P^2}{2 m_f}
-\sum_{q \neq 0} g_q \lk f_q+f_q^*\rk 
+\frac{1}{2 m_f} \lk \sum_{q \neq 0} q|f_q|^2\rk^2
\nn
&+&
\sum_{q \neq 0} \lk E_q-\frac{q \cdot P}{m_f}
+\frac{q^2}{2 m_f}\rk |f_q|^2 
+g_{bf} n_0.
\label{EVpol}
\eeq
The stationary equation ${\delta E_{pol}(P)}/{\delta f_q} =0$ 
determines the phonon momentum amplitude $f_q=f_{q;P}$
(we denote the $P$ dependence of the solution $f_{q;P}$ 
explicitly for later convenience): 
\beq
f_{q;P}=
-g_q \ldk E_q +\frac{q^2-2(1-\eta) q \cdot P}{2 m_f}\rdk^{-1},
\label{fqP}
\eeq 
where the drag parameter $\eta$ is determined from the self-consistency condition:
\beq
\eta P=\langle x;P| \sum_{q \neq 0} q C_q^\dagger C_q |x;P\rangle 
 =\sum_{q \neq 0} q |f_{q;P}|^2.
\label{Condeta}
\eeq
It implies that the mean value of the phonon momentum is proportional to
the polaron momentum.
% because only the polaron momentum gives an anisotropy of the system. 
Substituting (\ref{fqP}) into (\ref{EVpol}), 
we obtain the single polaron energy with the momentum $P$:
\beq
E_{pol}(P)
&=&
g_{bf} n_0
+\frac{(1-\eta^2) P^2}{2 m_f}
-g_{bf}^2 n_0 \int_q (u_q -v_q)^2 
\ldk E_q +\frac{q^2-2(1-\eta) q\cdot P}{2 m_f}\rdk^{-1}
\nn
&=&
g_{bf} n_0
-g_{bf}^2 n_0 \int_q
\frac{(u_q-v_q)^2}{E_q+\frac{q^2}{2 m_f}}
+\frac{P^2}{2 m_{eff}}  
 +\mathcal{O}(P^4), 
\label{epol}
\eeq
where the polaron effective mass $m_{eff}$ is defined by  
\beq
m_{eff}=\frac{m_f}{1-\eta}. 
\label{polaroneffmassX}
\eeq
The drag parameter $\eta$ is represented as 
$\eta = s/(1+s)$ using the parameter $s$ 
\footnote{see Appendix~\ref{ApCals} for derivation.}: 
\beq
s
=
\frac{32(1+R)^2}{3} 
\frac{a_{bf}^2 n_0^{\frac{1}{2}}}{a_{bb}^{\frac{1}{2}}}\int_0^\infty {\rm d}z 
\frac{z^2}
{\sqrt{z^2+16\pi}\lk
R\sqrt{z^2+16\pi}+z\rk^3}, 
\label{ExFs}
\eeq
where $R =m_f/m_b$ is the mass ratio. 
The formula (\ref{ExFs}) shows that 
the parameter $s$ is proportional to 
the so-called polaronic-coupling parameter 
$\alpha=\frac{a_{bf}^2}{\xi a_{bb}}\propto a_{bf}^2n_0^{1/2}/a_{bb}^{1/2}$, 
and that, at $\eta =1$ or equivalently $s\rightarrow \infty$,
the polaron effective mass $m_{eff}$ becomes infinite. 
The dependence of the parameter $s$ on the mass ratio $R$ is shown in Fig~\ref{test1}.
% 
%%%%%%%%%%%%%%%%%%%%%%%%%%%%%%%%%%%%%%%%%%%%%%%%%%%%%%%%%%%%%
%%%%%%%%%%%%%%%%%%%%%%%%%%%%%%%%%%%%%%%%%%%%%%%%%%%%%%%%%%%%%%
\subsection{Estimation of polaron size}
We also investigate the property of 
the phonon distribution around the polaron. 
Using the phonon field operator 
$\phi_{ph}(r)=V^{-1/2}\sum_p e^{ip\cdot r}C_p$, 
the photon-distribution probability at a position $r$ is given by 
$\langle x;P|\phi_{ph}^\dagger(r) \phi_{ph}(r) |x;P \rangle
= |f(r-x)|^2$, 
where $f(r)=V^{-1/2}\sum_p e^{ir\cdot p} f_{p;P}$ 
is the inverse Fourier transform of $f_{q;P}$. 
The phonon spatial amplitude $f(r)$ for the static polaron ($P=0$) is given by 
\beq
f(r)
&=& 
-\frac{a_{bf} n_0^{\frac{1}{2}} (R+1) }{\pi} \nn
& &\times \int_{-\infty}^\infty {\rm d}q \;
\frac{e^{i q r}}{ir} 
\lk \frac{q^2}{q^2 +\frac{2}{ \xi^2 }} \rk^{\frac{1}{4}}
\frac{q}{(R^2-1) q^2 +\frac{2R^2}{\xi^2}} 
\lk \frac{R\sqrt{q^2\lk q^2+\frac{2}{\xi^2}\rk}}{q^2}-1 \rk,  
\label{size1}
\nonumber
\eeq
where $\xi=1/\sqrt{8\pi n_0 a_{bb}}$ is the coherence length of the BEC. 
The detailed behavior of $f(r)$ depends on the structure of 
the momentum distribution $f_{q;0}$ given in (\ref{fqP}), 
which corresponds to the phonon propagator in the static limit. 
As seen from the above expression of $f(r)$, 
the integrand has the pole at $q=\pm \sqrt{\frac{2R^2}{1-R^2}}\xi^{-1}$ for $0 <R < 1$ 
but with the vanishing residue, 
thus in this case we can estimate by the dimensional analysis that $f(r)\sim r^{-5/2}$ for large $r$. 
While for $R > 1$, the pole emerges on the imaginary axis, at $q=\pm i \sqrt{\frac{2R^2}{R^2-1}}\xi^{-1}$, 
which implies that the size of the spatial distribution may be set by the order of the coherence length $\xi$. 
However, this situation is not exactly the same as 
that the coherence length sets the size of the Yukawa type potential between two probe fermions 
%$V(r)\sim e^{-\sqrt{2}r/\xi}/r$, 
(see Appendix~\ref{PotproF} for detail), 
since the residue theorem cannot be applied directly 
because of the existence of the branch points at $q=0, \pm i\sqrt{2}\xi^{-1}$ in this case. 
We postpone the rigorous discussion on the asymptotic analysis for the large $r$ behavior elsewhere. 

Nevertheless, such a mass-ratio dependence has also been observed 
in numerical simulations for the single-polaron for $R\le 1$ \cite{Grusdt1}; 
as the boson-fermion interaction varies from weak to strong couplings, 
non-perturbative approaches give a variety of results but show the deviation 
from mean-field approaches.
Thus, the qualitative change in the phonon spatial amplitude
shows the deviation from the mean-field to the non-perturbative regimes
which occur around $R \sim 1$.
Actually, the mean-field solution in the LLP theory 
becomes exact in the limit of
$m_f \rightarrow \infty$ ($R \rightarrow \infty$), 
where the effective phonon-phonon coupling vanishes. 

%%%%%%%%%%%%%%%%%%%%%%%%%%%%%%%%%%%%%%%%%%%%%%%%%%%%%%%
\subsection{Validity of Fr{\"o}hlich Hamiltonian }
%%%%%%%%%%%%%%%%%%%%%%%%%%%%%%%%%%%%%%%%%%%%%%%%%%%%%%%
In construction of the Fr{\"o}hlich type Hamiltonian, 
we have dropped the four-point interaction among excited phonons ($q\neq 0$) using the Bogoliubov approximation, 
which may be allowed when the number of excited phonons $N_{ph}=n_{ph}V$ is much small 
with respect to that of the condensed bosons ($q=0$): 
$n_{ph}/n_0 \ll 1$ as the density ratio. 
The number of bosons $N_b$ is expressed in terms of the phonon operators by 
\beq
N_b= N_0 + 
\sum_{q\neq 0} \ldk v_q^2 + \lk v_q^2+u_q^2\rk C_q^\dagger C_q 
-v_qu_q\lk C_qC_{-q}+C_{-q}^\dagger C_q^\dagger \rk \rdk, 
\eeq 
where the first term is the condensed bosons, 
the second term accounts for the virtual phonon excitations given by 
\beq
N_{ph}^v\equiv \sum_{q\neq 0} v_q^2 =N_0\frac{\lk n_0^{1/3} a_{bb}\rk^{3/2}}{3\sqrt{\pi}}, 
\eeq
and the real phonons excited by the boson-fermion coupling 
is evaluated for the LLP ground state to be, 
in the probe approximation ($m_f\rightarrow \infty $), 
\begin{equation} 
N_{ph}^r\equiv \sum_{q\neq 0} \langle  \lk v_q^2+u_q^2\rk C_q^\dagger C_q 
-v_qu_q\lk C_qC_{-q}+C_{-q}^\dagger C_q^\dagger \rk 
\rangle 
=V\int_q |f_{q;P}|^2
\simeq \frac{2 \lk n_0^{1/3}a_{bf}\rk^2}{\sqrt{\pi} \lk n_0^{1/3} a_{bb}\rk^{1/2}}, 
\end{equation} 
where we have used the relation: 
$\langle x;P|\int_r \phi_{ph}^\dagger(r) 
\phi_{ph}(r) |x;P \rangle= \sum_{q \neq 0} |f_{q;P}|^2$. 
The $N_{ph}^r$ corresponds to the excited phonons by the single fermion, 
and if the spatial extension of the phonon cloud is of the order of the coherence length, 
$\xi=1/\sqrt{8\pi a_{bb}n_0}$, 
for heavy fermions as discussed above 
and also from the momentum dependence of the Yukawa coupling \cite{Grusdt2,Bruderer1}, 
the density of the excited phonons around the fermion 
may be estimated by $n_{ph}^r\simeq N_{ph}^r \xi^{-3}$. 
As result, a condition for the Fr{\"o}hlich Hamiltonian is given by 
\beq
\frac{n_{ph}}{n_0}=\frac{n_{ph}^v+n_{ph}^r}{n_0}
=\frac{\lk n_0^{1/3} a_{bb}\rk^{3/2}}{3\sqrt{\pi}}
+2^{11/2}\pi \lk n_0^{1/3}a_{bf}\rk^2\lk n_0^{1/3} a_{bb}\rk \ll 1. 
\eeq
When only the second term is kept for a small boson-boson scattering length, 
the above formula is consistent with those obtained in \cite{Grusdt2,Bruderer1}. 

Now we define the density of excited phonons by a different way:  
$n_{ph}^r\equiv N_{ph}^rN_f/V=N_{ph}^r n_f$ being multiplied by the density of fermions, 
which is for the single fermion $N_f=1$ at moment, and can be used in the thermodynamic limit for many fermions. 
This leads to a condition: 
\beq
\frac{n_{ph}}{n_0}=\frac{n_{ph}^v+n_{ph}^r}{n_0}
=\frac{\lk n_0^{1/3} a_{bb}\rk^{3/2}}{3\sqrt{\pi}}
+\frac{2 \lk n_0^{1/3}a_{bf}\rk^2}{\sqrt{\pi} \lk n_0^{1/3} a_{bb}\rk^{1/2}}\frac{n_f}{n_0} \ll 1. 
\eeq
The above formula accounts for an averaged number of excited phonons per fermion, 
and thus may gives a validity condition of the Fr{\" o}hlich Hamiltonian for many-body polaron systems. 

%%%%%%%%%%%%%%%%%%%%%%%%%%%%%%%%%%%%%%%%%%%%%
\section{Polaron gas in BEC: Lemmens-Devreese-Brosens method}
%%%%%%%%%%%%%%%%%%%%%%%%%%%%%%%%%%%%%%%%%%%%%

In the previous section, 
we have discussed the single fermion immersed in the BEC, 
which behaves as a polaron dressed with Bogoliubov phonons.
Now we study the system of a dilute but finite density of fermions. 
If the interparticle distance of fermions 
is much larger than the size of each polaron,  
which is to be $\sim \xi$ for $R>1$, 
the dilute system of fermions should be described as the dilute polaron gas. 

In order to evaluate the ground state properties of such a gas, 
we will first employ the method by Lemmens-Devreese-Brosens (LDB)
originally developed for the electron-polaron gas \cite{LDB1}, 
and, then
we will propose a more general method which incorporates the drag effect as in the LLP 
in the next section. 
Both methods are based on the second quantization of the LLP theory for many-fermion systems. 
As shown in Eq.~(\ref{LLPgs}) 
the unitary transformation of the Hamiltonian in the LLP theory is composed of two consecutive transformations 
$S$ and $T$ defined in (\ref{transS}) and (\ref{transT}):
$H \rightarrow T^{-1}S^{-1}HST$. 
These two transformations are not commutable, 
and the another transformation $U$ has been introduced in (\ref{transU}).
In the calculation of the expectation value by the phonon vacuum in (\ref{EVpol}), 
we could use the $U$ transformation only 
because the phonon vacuum is invariant 
against the $S$ transformation: $S\left| 0 \rangle \right.= \left| 0 \rangle\right.$.
It means that the $U$ transformation absorbs the effects of phonon, 
and plays a role of making the fermions dressed with the phonon cloud.  
Thus we will eventually construct the second-quantized $U$ transformations, 
and use them for describing the polaron gas.
\footnote{
Note that these methods for polaron systems trace back 
to the scalar meson theory \cite{Schweber}, 
and also to the nucleon with meson cloud by Tomonaga \cite{Tomonaga}. }

%%%%%%%%%%%%%%%%%%%%%%%%%%%%%%%
\subsection{LDB transformation} 
%%%%%%%%%%%%%%%%%%%%%%%%%%%%%%%
In Lemmens-Devreese-Brosens (LDB) theory, 
the transformation $U=e^{Q(r)}$ in LLP is extended to
\beq
Q(r) \rightarrow \sum_i Q(r_i) 
=\int_r \hat{n}_f(r) Q(r),
\nonumber
\eeq
where $r_i$ is the position of $i$-th polaron,
the fermion density operator $\hat{n}_f(r)$ is defined by $\hat{n}_f(r)=\psi^\dagger(r) \psi(r)$. 
The boson operator $Q(r)$ is the same as that in (\ref{transU}). 
Thus the $U$ transformation of LDB becomes\footnote{
Our notation is consistent with 
the original LDB transformation  
up to the definition $f_q \rightarrow -f_q^*$.}
\beq
U
&=& 
\exp\ldk \sum_{q,P}  
\lk f_q a_{P-q}^\dagger a_P C_q^\dagger-
f_{q}^* a_{P+q}^\dagger a_P C_q\rk \rdk.  
\label{LDBtra1}
\eeq
Note that no $P$-dependence is assumed for the phonon momentum amplitude $ f_q $ in LDB. 
The anti-Hermiticity $Q^\dagger(r)=-Q(r)$ guarantees the unitary condition  
$U^\dagger U=1$. 

In the $ U $ transformation of the LDB method, 
no momentum-anisotropy is assumed in the momentum amplitude $f_q$, i.e., $f_q^*=f_{-q}$, 
from which we can prove $\sum_{q \neq 0} q |f_q|^2 =0$;
in comparison with (\ref{Condeta}),
it shows that no drag effect is included in the LDB formulation ($\eta =0$).  
Inclusion of the anisotropic effect $f_q^*\neq f_{-q}$ is presented in Appendix \ref{ApLDBaI}. 

The transformation laws with the $U$ transformation (\ref{LDBtra1})
become 
\beq
U^{-1} \psi(x) U 
&=&
e^{Q(x)} \psi(x),  
\nn
U^{-1}\psi^\dagger(x) U
&=&
\psi^\dagger(x)e^{-Q(x)}, 
%%%%%%%%%%%%%%
\nn
U^{-1}\nabla\psi(x)U
&=&\nabla(U^{-1}\psi(x)U)
=
e^{Q(x)} 
\ldk \nabla+\nabla Q(x)\rdk 
\psi(x), 
%%%%%%%%%%%%%%%%%%%%%%%%%%%
\nn
U^{-1} \nabla\psi^\dagger(x) U
&=&\nabla(U^{-1} \psi^\dagger(x) U)
=
\ldk 
\nabla\psi^\dagger(x)
-\psi^\dagger(x)\nabla Q(x)\rdk  
e^{-Q(x)} 
\nonumber
\eeq
for fermion fields and their derivatives,
and 
\beq 
U^{-1} C_q U
&=&
C_q + f_q \int_r \hat{n}_f(r) e^{-ir\cdot q}, 
\nn
U^{-1} C_q^\dagger U
&=&
C_q^\dagger + f_q^* \int_r \hat{n}_f(r) e^{ir\cdot q} 
\nonumber
\eeq
for the phonon fields. 
Note that 
the transformed fermion field operators have the factor $e^{Q(x)}$, 
which entails the phonon cloud. 

Thus the Hamiltonian (\ref{FrH}) is transformed as
\beq 
U^{-1}HU
&=&
U^{-1}H_fU
+U^{-1}H_{b}U
+U^{-1}H_{int}U
\eeq
where
\beq
U^{-1} H_f U
&=&
\frac{1}{2 m_f} 
\int_x 
\ldk 
\nabla\psi^\dagger(x)
-\psi^\dagger(x)\nabla Q(x)
\rdk
\cdot
\ldk 
\nabla\psi(x)
+\nabla Q(x)\psi(x)
\rdk, 
%%%%%%%%%%%%%%%%%%%
\\
U^{-1}H_{b}U 
&=&\sum_{k} E_k 
\ldk C_k^\dagger + f_k^* \int_r \hat{n}_f(r) e^{ir\cdot k}\rdk 
\ldk C_k + f_k \int_r \hat{n}_f(r) e^{-ir\cdot k} \rdk,  
%%%%%%%%%%%%%%%%%%%
\\
U^{-1}H_{int}U
&=&
\int_r \hat{n}_f(r) 
\sum_{q \neq 0}
g_q
\Big[ e^{-i q \cdot r} C_q^\dagger +e^{i q \cdot r} C_q \nn
& &\qquad\qquad\qquad\quad
   + \int_x \hat{n}_f(x) 
   \left\{  f_q^* e^{i q \cdot (x-r)} 
         +f_q e^{-i q \cdot (x-r)} \right\} \Big]
+g_{bf} n_0 N_f. 
\eeq
Taking the normal ordering for phonon fields, 
we classify the terms of the Hamiltonian 
in the order of fermion fields:
\beq
U^{-1}HU
&=& H^{(mf)}+H^{(2)}+H^{(4)}+H^{(no)}, 
\label{tfH1}
\eeq
where the first term is the mean-field contribution 
\beq
H^{(mf)}
&=& g_{bf} n_0 N_f, \label{mfpart}
%%%%%%%%%%%%%%%%%%%%%%%%%%%
\\
H^{(2)}
&=&
\int_x \psi^\dagger(x) 
\ldk 
-\frac{\nabla^2 }{2 m_f}
+\sum_{q \neq 0} \frac{q^2}{2 m_f} |f_q|^2 
\rdk 
\psi(x), 
%%%%%%%%%%%%%%%%%%%%%%%%%%%
\\
H^{(4)}
&=&
\int_x \int_y 
\hat{n}_f(x) \hat{n}_f(y)
\sum_{q \neq 0} 
e^{i q \cdot (x-y)}
\ldk E_q |f_q|^2 + g_q \lk f_{-q}^* +f_q \rk \rdk, 
\label{4f1} 
\eeq
and 
the $H^{(no)}$ includes the normal ordered products of phonon fields 
such as $C_q^\dagger C_q$, $C_qC_q$, and $C_q^\dagger C_q^\dagger$, 
which will vanish in the expectation value for the phonon vacuum state.  
%
%%%%%%%%%%%%%%%%%%%%%%%%%%
\subsection{Ground state energy} 
%%%%%%%%%%%%%%%%%%%%%%%%%%
In this section,
we evaluate the ground state energy $E$
from the expectation value of the transformed Hamiltonian (\ref{tfH1}) 
with the variational ground state of the polaron gas 
that is constructed as the product state of the phonon vacuum 
and the many-fermion state.
When the phonon vacuum is operated on the Hamiltonian, 
the phonon normal-ordered term $H^{(no)}$ vanishes 
and the other terms including fermion fields remain. 
Then, we obtain the energy expectation value per fermion:
\beq
\frac{E}{N}_f=
E_{kin}
+g_{bf}n_0
+\frac{1}{2 m_f}
\sum_{q \neq 0} q^2 |f_q|^2 
+
\sum_{q \neq 0} 
S(q)
\left\{ E_q|f_q|^2 
+
g_q \lk f_{-q}^*+f_q \rk 
\right\},  
\label{EpF}
\eeq
where the kinetic energy per fermion is defined by 
\beq
E_{kin}
&=&
-\frac{N_f^{-1}}{2 m_f} \int_x 
\langle
\psi^\dagger(x)
\nabla^2 \psi(x)
\rangle, 
\eeq
where $\langle \cdots \rangle$ denotes the expectation value by the many-fermion state, 
and the structure factor $S(q)$  in the last term is defined by 
\beq
S(q)=
\frac{1}{N_f} \int_r \int_x 
e^{i(r-x) \cdot q}
\langle 
\hat{n}_f(r) \hat{n}_f(x)
\rangle,
\eeq
which encodes the fermion contribution in the interaction energy of (\ref{EpF}).

The stationary condition ${\delta E}/{\delta f_q^*}=0$
determines the momentum amplitude:
\beq
f_q=
-\frac{g_q}{E_q+\frac{q^2}{2 m_f S(q)}}.
\label{fq1}
\eeq
Substituting it into (\ref{EpF}), 
we obtain the ground state energy of the polaron gas in LDB:
\beq
\frac{E}{N_f}
&=& 
E_{kin} - 
g_{bf}^2 n_0 \int_q
\frac{S(q)(u_q-v_q)^2}{E_q+\frac{q^2}{2 m_f S(q)} }
+g_{bf}n_0. 
\label{recoil1}
\eeq

Using the Hartree-Fock approximation \citep{Fetter-Walecka} for the many-fermion state, 
the structure factor $S(q)$ is given by 
\beq
S(q)
&=&
\frac{1}{N_f} \sum_{k,p}  
\langle a_{k+q}^\dagger a_{k} a_{p-q}^\dagger
a_{p}\rangle
\simeq
-\frac{1}{n_f}
\int_k \theta(q_F-|k+q|)\theta(q_F-|k|)+1
\nn
&=&
\ltk \begin{array}{cc}
\frac{3}{2}\frac{q}{2 q_F}
-\frac{1}{2}\lk \frac{q}{2 q_F}\rk^3 & 
\mbox{ for } q<2 q_F \\
1 & \mbox{ for } q\ge 2 q_F
\end{array}
\right.
\eeq
where $q_F =(6\pi^2 n_f)^{1/3}$ is the Fermi momentum corresponding to the fermion density $n_f =N_f/V$.
Also, in this approximation, 
the kinetic energy par fermion is given by 
\beq
E_{kin}=\frac{3}{5}\epsilon_F, 
\eeq
where $\epsilon_F=q_F^2/2m_f$ is the Fermi energy. 

\subsection{Renormalization of boson-fermion interaction}

The interaction energy in (\ref{recoil1}) has an ultra-violet (UV) divergence,
which is attributed to the microscopic behavior 
of the low energy $s$-wave scattering amplitude, 
and can be renormalized in terms of 
$s$-wave scattering length $a_{bf}$ observable in experiments \cite{Grusdt1}.
From Eq.~(\ref{Tmatrix}) in the T-matrix approximation,
the coupling constant $g_{bf}$ is represented in terms of the scattering length $a_{bf}$
at the low energy limit: 
\beq
g_{bf}
= 
\frac{2\pi a_{bf}}{m_{bf}}
\left\{ 1+\frac{2\pi a_{bf}}{m_{bf}}
\int_q \frac{1}{q^2/2 m_{bf}}
+\cdots \right\},
\label{gbfT}
\eeq
where the divergent integral is regularized by the UV cutoff $\sim r_0^{-1}$.
It should be noted that Eq.~(\ref{gbfT}) is valid in the case of the weak boson-fermion interaction
(small value of $a_{bf}/r_0$).

The leading divergence of the interaction integral in (\ref{recoil1}) can be extracted as 
\beq
-g_{bf}^2 n_0 \int_q
\frac{S^2(q)}
{S(q)\lk \epsilon_q+2 g_{bb}n_0\rk+R^{-1}E_q}
&=&
-g_{bf}^2 n_0 \int_q
\frac{1}
{\lk R^{-1}+1\rk \epsilon_q} +\cdots \nn
&\simeq& 
-\lk \frac{2\pi a_{bf}}{m_{bf}} \rk^2 n_0 \int_q
\frac{1}
{q^2/2 m_{bf}} +\cdots, 
\label{EintDiv}
\eeq
where $R=m_f/m_b $ is the boson-fermion mass ratio.
Using Eq.~(\ref{gbfT}), 
the leading-order contribution of the mean-field energy $g_{bf} n_0$ in (\ref{recoil1}) 
becomes 
\beq
g_{bf}n_0 \simeq 
\frac{2\pi a_{bf}}{m_{bf}}n_0
+\lk \frac{2\pi a_{bf}}{m_{bf}}\rk^2
n_0 \int_q\frac{1}{q^2/2 m_{bf}}, 
\label{mfCont}
\eeq
the second term of which exactly cancels out the divergent term in (\ref{EintDiv}). 

Finally, the renormalized ground state energy thus becomes 
\beq
\frac{E}{N_f}
&=&
\frac{2\pi a_{bf}}{m_{bf}}n_0 
+\lk \frac{2\pi a_{bf}}{m_{bf}} \rk^2 n_0 
\int_q \frac{1}{q^2/2 m_{bf}}
\nn 
&+&
E_{kin}
-\lk \frac{2\pi a_{bf}}{m_{bf}} \rk^2 n_0 \int_q
\frac{S^2(q)}
{S(q) \lk \epsilon_q +2 g_{bb} n_0 \rk +\frac{m_b}{m_f} E_q}, 
\label{ELDB1}
\eeq
which is consistent with that obtained in \cite{LDB2} 
in the absence of the bare fermion-fermion interaction.
%%%%%%%%%%%%%%%%%%%%%%%%%%%%%%%%%%%%%%%%%%%%%
\section{Polaron gas in BEC: a many-body extension of LLP} 
%%%%%%%%%%%%%%%%%%%%%%%%%%%%%%%%%%%%%%%%%%%%%
In the LDB method,
no drag effect is included in the phonon cloud around polarons ($\eta=0$).  
In the case of the dilute fermion gas, 
it is natural to expect that the fermions undergo such an effect 
as in the single-polaron LLP. 
In order to incorporate the drag effect in the many-polaron system,  
we use the extended $U$ transformation 
\beq
U =e^S =
\exp\ldk \sum_{q,P}  
\lk f_{q;P} a_{P-q}^\dagger a_P C_q^\dagger
-f_{q;P}^* a_{P}^\dagger a_{P-q} C_{q} \rk \rdk, 
\label{eLLPUni}
\eeq
where the $P$-dependent phonon momentum amplitude $f_{q;P}$ is used 
instead of $f_q$ in (\ref{LDBtra1}).  
Noted that the above $U$ transformation keeps the unitarity condition $ U^\dagger U=1 $, 
and includes the $U$ transformation of the LDB method 
as a special case where no $P$-dependence exists in the function $f_{q;P}$.
Although the way to extend the single-polaron LLP to many-fermion systems is not the unique, 
it seems very reasonable to develop the method with the transformation (\ref{eLLPUni})
because of the success of LLP and LDB; 
thus, we take the extended $U$ transformation to include the drug effect and 
call the method the extended LLP (eLLP).
In what follows 
we assume that the $f_{q;P}$ in (\ref{eLLPUni}) is a real function as in LLP and LDB. 

%%%%%%%%%%%%%%%%%%%%%%%%%%%%%
\subsection{Transformations of field operators}
%%%%%%%%%%%%%%%%%%%%%%%%%%%%%
The existence of the drag effect ($\eta \neq 0$) of the photon cloud around polarons 
means that the cloud shares the part of the total polaron momentum 
as in the single-polaron LLP through the relation: 
$\eta P=\sum_{q \neq 0} q |f_{q;P}|^2$. 
Then, because of the finite value of $\eta$, 
the function $f_{q;P} $ should have momentum anisotropy and be expanded as 
$f_{q;P} = c_0 + c_1 \lk q\cdot P\rk + c_2 \lk q\cdot P\rk^2 + \cdots$. 
Since the momentum $P$ is of the order of the Fermi momentum at most,  
only the leading term dominates in the dilute regime of fermions. 
With this observation, we consider the transformation of the field operators by (\ref{eLLPUni}).

Using the general formula of the similarity transformation,
the fermion field $\psi(x)$ obeys the transformation:
\beq
U^{-1} \psi(x )U =\psi(x) 
+\ldk -S, \psi(x) \rdk 
+\frac{1}{2!}
\ldk -S, \ldk -S, \psi(x) \rdk \rdk +\cdots, 
\label{Unew}
\eeq
where the commutators are given by 
\beq
\ldk -S, \psi(x) \rdk 
&=& \int_y \alpha(x,y) \psi(y), 
\nn
\ldk -S,\ldk -S, \psi(x) \rdk \rdk 
&=& \int_{y,z}\ldk w(x,y) \delta(y-z)+\alpha(x,y)\alpha(y,z)\rdk \psi(z), 
\quad \hbox{and so on},
\eeq 
where operators $\alpha(x,y)$ and $w(x,y)$ are defined as
\beq
\alpha(x,y)
&=& 
\frac{1}{V} \sum_{k,Q} e^{i (Q-k) x} e^{-i Q y}
\left\{ f_{k;Q}  C_{k}^\dagger 
-f_{-k;Q-k} C_{-k}\right\}, 
\label{alphaxy}
\\
w(x,y)
&=& 
\frac{1}{V} \sum_{k,P,Q} e^{i(Q-k)x}e^{-i Q y}
\left\{ f_{k;Q} f_{k;P}  
-f_{-k;Q-k} f_{-k;P-k}\right\} a_P^\dagger a_{P-k}. 
\label{wxy}
\eeq
We note that the operators $\alpha(x,y)$ and $w(x',y')$ are computable:
$\ldk \alpha(x,y), w(x',y')\rdk =0$. 
The transformation (\ref{Unew}) generates infinitely many terms of 
higher order products of phonon and fermion fields, 
which are expected to be less contributed in dilute Fermi gas. 
However, the terms producing the drag effect are summed up to be an 
exponential form:
\beq
U^{-1} a_P U \simeq \frac{1}{V} \sum_Q \int_{x,y} e^{-i P \cdot x+i Q \cdot y} 
\langle x| e^{\hat{A}} |y\rangle a_Q,  %+ \cdots 
\label{fertra1}
\eeq
where we have used the bracket notation:
\beq 
\langle x| e^{\hat{A}} | y \rangle &=&
\langle x | y \rangle +\langle x | \hat{A} | y \rangle 
+\frac{1}{2!} 
\int_z \langle x | \hat{A} | z \rangle 
\langle z | \hat{A} | y \rangle 
+\cdots, 
\nn 
&=& 
\delta(x-y)
+A(x,y) 
+\frac{1}{2!} \int_z A(x,z) A(z,y) 
+\cdots
\eeq
and the operator $A(x,y)$ is defined as
\beq
     A(x,y) =\langle x |A| y \rangle 
            =\alpha(x,y)+\frac{1}{2} w(x,y). 
\eeq
The derivation of the above equations is given in Appendix~\ref{eLLPtra}. 

In contrast, 
the transformation of the phonon field is obtained in the exact form:
\beq
U^{-1} C_q U=
C_q +\sum_P f_{q;P} a_{P-q}^\dagger a_P, 
\label{phontra1}\\
U^{-1} C^\dagger_q U
=C^\dagger_q +\sum_P f_{q;P} a_P^\dagger a_{P-k}. 
\label{phontra2}
\eeq 
Here we should note the following properties:  
The transformation (\ref{eLLPUni}) preserves the total momentum of the system, 
and the approximate transformation of fermion (\ref{fertra1}) 
together with exact ones (\ref{phontra1},\ref{phontra2}) also gives 
the exact transformation for the total momentum operator, 
\beq
 \hat{P}
 =\sum_P P a^\dagger_P a_P +\sum_q q C^\dagger_q C_q, 
\eeq
i.e., commutes with it, $[U,\hat{P}]=0$ (see Appendix \ref{ApMom} in detail), 
which supports the use of (\ref{fertra1}) in the present calculations of eLLP. 
In addition, as discussed in the next section, 
the transformation (\ref{eLLPUni}) provides exactly 
the same results of the LLP theory for the single fermion state, 
and thus a natural many-body extension of the LLP theory. 

%%%%%%%%%%%%%%%%%%%%
\subsection{Transformation of Hamiltonian}
%%%%%%%%%%%%%%%%%%%%
Under the transformations in (\ref{fertra1}) and (\ref{phontra1}), 
the Fr\"ohlich Hamiltonian (\ref{FrH}) becomes 
\beq
U^{-1} H U \simeq H^{(mf)}+H_F' +H_B' +H_{I}'.
\label{FrHUnew}
\eeq 
The first term $H^{(mf)}=g_{bf} n_0 N_f$ is the mean-field contribution, 
and $H_F'$, $H_B'$, and $H_I'$ are represented by
\beq
H_F' &=&
\frac{1}{2 m_f} \sum_P a_P^\dagger ( P -u_P)^2 a_P
-\frac{1}{2 m_f}\sum_{k,P} (2 P -k -u_P-u_{P-k}) \cdot k 
a_{P-k}^\dagger X_{k,P} a_P
\nn
&&+\frac{1}{2 m_f} \sum_{k,k',P} k \cdot k'  
a_{P-k}^\dagger X_{k,P} X_{k',P}^\dagger a_{P-k'}, 
\\
H_B' &=& 
\sum_{k} E_k 
\lk C_k^\dagger +\sum_P f_{k;P} a_P^\dagger a_{P-k} \rk 
\lk C_k+\sum_P f_{k;P} a_{P-k}^\dagger a_{P} \rk, 
\\
H_{I}' &=& 
\sum_{k,P,Q} g_k \lk C_k^\dagger +\sum_P f_{k;P} a_{P}^\dagger a_{P-k} \rk
a_{Q-k}^\dagger a_Q
+\hbox{h.c.}
\eeq
where 
$u_P =\sum_k k f_{k;P}^2$ and
\beq
X_{k,Q} &=& \frac{1}{V} \int_{x,y} e^{-i (Q-k) x} e^{i Q y} A(x,y) 
\nn
&=&f_{k;Q} C_k^\dagger -f_{-k;Q-k} C_{-k} 
+\frac{1}{2} \sum_P 
\left\{ 
f_{k;Q} f_{k;P} -f_{-k;k-Q} f_{-k;P-k}
\right\} 
a_P^\dagger a_{P-k}. 
\eeq

After the normal ordering operation for the phonon field operators, 
we rearrange the Hamiltonian (\ref{FrHUnew}) 
in the order of fermion field operators as 
\beq
U^{-1}HU 
&\simeq& H^{(mf)} + H^{'(2)} +  H^{'(4)}+H'^{(no)},  
\label{Hfermion}
\eeq
where the momentum representation of $H^{'(2)}$ and $H^{'(4)}$ are given by 
\beq
H^{'(2)}
&=&
\sum_{P} 
\ldk 
\frac{(P-u_P)^2 }{2 m_f}
+\sum_{q \neq 0} \left( E_q +\frac{q^2}{2 m_f} \right) f_{q;P}^2 
-2\sum_{q \neq 0} g_q  f_{q;P} 
\right\}
a_P^\dagger a_P, 
\\
H^{'(4)}
&=&
-
\sum_{q,P,Q}
\left\{ E_q f_{q;P}f_{q;Q}+g_q ( f_{q;P}+f_{q;Q} ) \right\} 
a_P^\dagger a_{Q-q}^\dagger a_{P-q}a_Q,  
\label{4f2} 
\eeq
and the $H'^{(no)}$ including the phonon operators in the normal ordering 
vanishes when the expectation value is taken with the phonon vacuum. 
Note that we have dropped four- and six-body interactions generated from the fermion's kinetic term 
approximately for the dilute gas of our interest.  

%%%%%%%%%%%%%%%%%%%%
\subsection{Many-polaron ground state including drag effect}
%%%%%%%%%%%%%%%%%%%%
Now we calculate the expectation value of the Hamiltonian (\ref{Hfermion})
with the phonon vacuum and the many-fermion state.
Using the Hartree-Fock approximation for fermions as in the LDB method:
\beq
a_P^\dagger a_{Q-q}^\dagger a_{P-q}a_Q
\simeq 
a_{P}^\dagger a_{Q} 
\langle a_{Q-q}^\dagger a_{P-q} \rangle 
= 
a_{P}^\dagger a_{P} 
\theta\lk q_F-|P-q|\rk \delta_{P,Q}, 
\eeq
we obtain the energy expectation value:
\beq
E = g_{bf}n_0N_f+ 
\sum_{P} 
\theta(q_F-|P|)  \left\{ E_{pol}(P) - \epsilon^{F}(P) \right\}, 
\label{EExPresent}
\eeq
where the Fock exchange contribution $\epsilon^{F}(P)$ is given by 
\beq
\epsilon^{F}(P) =
\sum_{q \neq 0}
\theta\lk q_F-|P-q|\rk 
\left\{ E_q f_{q;P}^2 
+2 g_q f_{q;P} 
\right\}. 
\eeq

The stationary equation of the energy expectation value ${\delta E}/{\delta f_{q;P}^*} =0$ becomes 
\beq
\theta(q_F-|P|)
\ldk 
\frac{1}{m_f}
\lk \sum_k k f_{k;P}^2 \rk \cdot q f_{q;P}
+\lk \frac{q^2}{2 m_f}
-\frac{q\cdot P}{m_f}\rk f_{q;P} 
\right. 
\nn
\left. 
+
\ltk 1-\theta\lk q_F-|P-q|\rk \rtk
\lk 
E_q f_{q;P}+ g_{q}
\rk 
\rdk =0. 
\label{StEqpresent}
\eeq
Existence of the drag parameter $\eta$ shows the inclusion of the drag effect of the phonon cloud 
through the relation
$\eta P=\sum_k k|f_{k;P}|^2$ as in the single-polaron LLP,
with which we can solve the stationary equation (\ref{StEqpresent}) for $f_{q;P}$ ($|P|\le q_F$): 
\beq
f_{q;P}
&=&
-\frac{g_q}
{E_q+\frac{q^2}{2 m_f}
-\frac{q\cdot P(1-\eta)}{m_f}}
\theta\lk |P-q|-q_F\rk. 
\label{fq2}
\eeq
Substituting it into (\ref{EExPresent}) and using the variational condition, 
we obtain the ground-state energy: 
\beq
E
&=& g_{bf}n_0 N_f+
V\int_P \theta(q_F-|P|)E_f(P), 
\label{inte1}
\eeq
where the single-particle energy $E_f(P)$ is given by 
\beq
E_f(P)= 
\frac{1-\eta^2}{2 m_f}P^2
-
g_{bf}^2 n_0 
\int_q 
\theta\lk |P-q|-q_F\rk 
\frac{\lk u_q-v_q\rk^2} 
{E_q+\frac{q^2}{2 m_f}
-\frac{q\cdot P(1-\eta)}{m_f}}. 
\eeq

The interaction energy in the above expression includes the divergent term 
as appeared in LDB (section IV-C), 
hence we apply the same renormalization procedure 
in terms of the boson-fermion scattering length.
Finally, we obtain the renormalized ground state energy per fermion 
in the present formalism: 
\beq
\frac{E}{N_f}
&=&
\frac{2\pi a_{bf}}{m_{bf}}n_0 +\lk \frac{2\pi
a_{bf}}{m_{bf}}\rk^2 n_0 \int_q\frac{1}{q^2/2 m_{bf}}
+E_{kin}\lk 1-\eta^2 \rk
\nn 
&-&
\lk \frac{2\pi a_{bf}}{m_{bf}} \rk^2 \frac{n_0}{n_f} 
\int_{q,P}\theta\lk q_F-|P|\rk 
\theta\lk |P-q|-q_F\rk 
\frac{\lk u_q-v_q\rk^2} 
{E_q+\frac{q^2}{2 m_f}
-\frac{q\cdot P(1-\eta)}{m_f}}. 
\label{EOUR1}
\eeq
%

%%%%%%%%%%%%%%%%%%%%%%%%%%%%%%%%%%%%%%%%
\subsection{In-medium effective mass}
%%%%%%%%%%%%%%%%%%%%%%%%%%%%%%%%%%%%%%%%
Here we discuss the drag effect in the polaron gas, 
which manifests itself as a finite value of the parameter $\eta$. 
This parameter is determined from the self-consistent equation:  
\beq
\eta P
&=&\int_q q \frac{g_{bf}^2 n_0 (u_q-v_q)^2}
{\left( E_q+\frac{q^2}{2 m_f}-\frac{q\cdot P
(1-\eta)}{m_f} \right)^2}
\theta(|P-q|-q_F). 
\label{imeta}
\eeq
Expanding the right hand side to the order of $P$, 
we obtain 
\beq
\eta=\frac{X+W}{1+W}
\label{eta1}
\eeq
where
\beq
X
=
-\frac{2 g_{bf}^2 n_0}{3(2\pi)^2} 
\frac{\frac{q_F^5}{2 m_b}}
{E_{q_F}\lk E_{q_F}+\frac{q_F^2}{2 m_f}\rk^2}, 
\mbox{ and } 
W
=
g_{bf}^2 n_0 \int_q  
\frac{2\frac{q^2}{2 m_b}\frac{\lk q\cdot \hat{P}\rk^2}{m_f}}
{E_q\lk E_q+\frac{q^2}{2 m_f}\rk^3}
\theta(|q|-q_F) 
\eeq
with $\hat{P}=P/|P|$.

Using these results, 
we can expand the single particle energy of the fermion as 
\beq
E_f(P)&=&
-g_{bf}^2 n_0 
\int_q 
\theta\lk |q|-q_F\rk
\frac{\lk u_q-v_q\rk^2} 
{E_q+\frac{q^2}{2 m_f}}
+\frac{P^2}{2 m^*}
+\mathcal{O}\lk P^4 \rk, 
\eeq
and the effective mass $m^*$ reads 
\beq
\frac{m_f}{m^*}=1-\eta 
-\lk1-\eta -\frac{1-R^2}{2}\rk X-Z, 
\label{eLLPmass}
\eeq
where 
\beq
Z
&=&
\frac{g_{bf}^2n_0R^2}{3(2\pi)^2}
\frac{\frac{q_F^5}{2m_b}}{E_{q_F}^3}. 
\eeq
Note that  
these results reproduce the LLP theory for the single polaron 
in the dilute limit ($q_F\rightarrow 0$). 

For comparison with the LDB result, 
we employ the inertial mass $m^{\rm (in)}$, 
which corresponds to a linear response to the external velocity $v$ 
coupled with the total momentum operator of the whole system. 
The expression of the inertial mass in the LDB method 
for the system same as ours is given by Eq.~(15) in \cite{LDB2}: 
\beq
m^{\rm (in)}
=m_f+\frac{2}{3}\sum_q \frac{g_q^2 S(q)^2k^2}{\ldk E_q S(q)+q^2/2m_f\rdk^3}. 
\label{LDBmass}
\eeq
It is obvious from the behavior of $S(q)$ that 
the above expression reduces to the effective mass in the LLP theory 
at the low density limit. 

%%%%%%%%%%%%%%%%%%%%%%%%%%%%%%%%%%%%%%%%%%%%%
\section{Numerical results and discussion}
%%%%%%%%%%%%%%%%%%%%%%%%%%%%%%%%%%%%%%%%%%%%%
In this section, 
we present numerical results for the polaron gas in the BEC 
calculated in LDB and eLLP including the drag effect, 
and shortly explain how in-medium modifications of polaronic properties 
such as binding energy and effective mass can be observed in experiments. 
Also, we compare the present results with the 2nd order perturbation theory 
to clarify relations among these methods, 
and then briefly discuss a criterion for the validity 
of the mean-field type approximations used in this study. 
\subsection{Ground-state energy}
\begin{figure}[h]
  \begin{center}
    \begin{tabular}{ccc}
\resizebox{73mm}{!}{\includegraphics{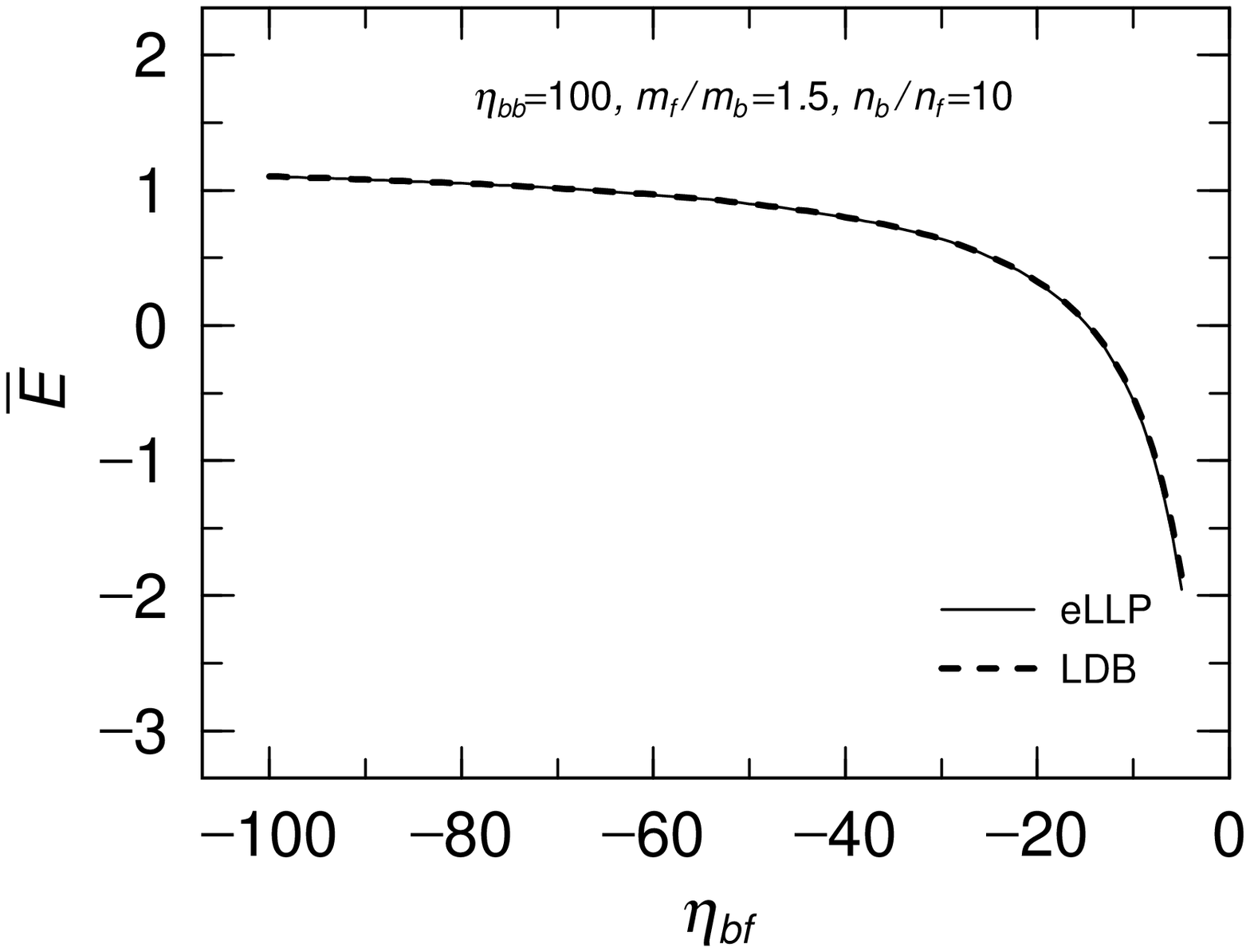}}& \hspace{0.5cm} &
\resizebox{80mm}{!}{\includegraphics{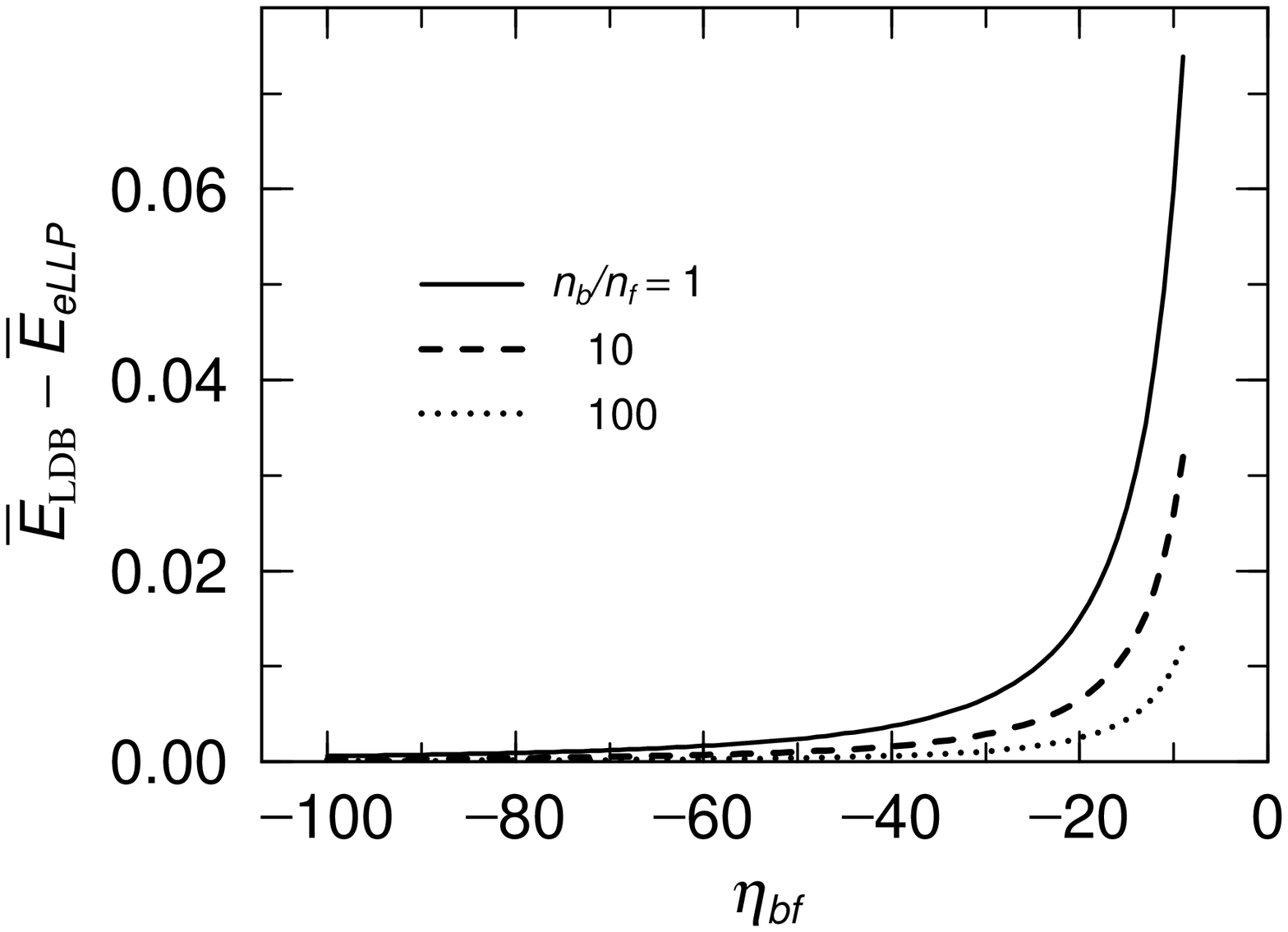}} \\
    \end{tabular}
    \caption{The ground state energies, 
    (\ref{ELDB1}) and (\ref{EOUR1}), in the eLLP and the LDB methods (left), 
    and their difference (right) as functions of $\eta_{bf}$. 
    Parameters are set as $\eta_{bb}=100$, $m_f/m_b=1.5 $, and  
    $n_b/n_f =1,10,100$.} 
    \label{FigOur1}
  \end{center}
\end{figure}

In Fig.~\ref{FigOur1}, 
we show 
the ground state energies calculated with (\ref{ELDB1}) and (\ref{EOUR1}) respectively in LDB and eLLP methods 
for the inverse of the boson-fermion scattering length;
the dimensionless energy $\bar{E}$ per fermion, 
the inverse of the scattering lengths, $\eta_{bf}$ and $\eta_{bb}$, are scaled as 
\beq
\bar{E}&=&\frac{E}{N_f E_0}, \quad
%\quad \mbox{and} \quad 
\eta_{bf}=\frac{1}{a_{bf}n_0^{1/3}}, \quad 
\eta_{bb}=\frac{1}{a_{bb}n_0^{1/3}}, 
\label{scaling}
\eeq 
where $E_0 =\frac{n_0^{2/3}}{2m_b}$ is the boson zero-point energy.
The boson-fermion mass ratio $R=m_f/m_b$ is fixed to $R=1.5$, 
and we approximate that bosons are all condensed, i.e., $n_b=n_0$.
For comparison with experimental setups, 
we refer to the system of ytterbium isotopes,  
$^{170}$Yb-$^{173}$Yb, with scattering lengths 
$a_{bb}=3.435$ nm and $a_{bf}=-4.373$ nm \cite{Enomoto1}, 
which are of the order of the atomic size so that the system is not strongly correlated. 
The spatial extension of the condensation estimated from a trap frequency 
($\omega\sim 2\pi\times10^2$ Hz) 
is of the order of $1\sim 10$ $\mu$m, 
and the number of the condensed bosons 
can be varied from $\sim 10^3$ to $\sim 10^6$, 
which then amount to $\eta_{bb}\sim 2.9\times 10^{1-2}$ 
and $\eta_{bf}\sim -2.3\times 10^{1-2}$. 
In what follows, we plot figures 
for values of the coupling strength up to $\eta_{bf}\sim -10$, 
which is consistent with the present approximations 
for weak/intermediate coupling regimes. 
The strongly correlated regime around the unitary limit, 
where strong correlation effects dominate, corresponds to roughly 
$|\eta_{bf}| \ll 10$ as illustrated in \cite{Rath1,Ardila2}. 

Fig.~\ref{FigOur1} (right) shows the energy difference between 
the eLLP and LDB ground states 
calculated for the boson-fermion mass ratios $Y =n_b/n_f =1,10,100$.
%
%%%%%%%%%%%%%%%%%%%%%%%%%%%%%%%%%%%%%%%%%%%%%%%%%%%%%%%%%%%%%
%%%%%%%%%%%%%%%%%%%%%%%%%%%%%%%%%%%%%%%%%%%%%%%%%%%%%%%%%%%%%%
We find that the ground state energies calculated in these two methods are almost on the top of each other;
however, detailed observation shows that eLLP gives slightly lower values of the ground-state energy 
and the difference becomes larger in the case of lower fermion densities. 
Possible reasons for these results are that the $P$-dependence of $f_{q;P}$ extends the variational space in the eLLP method, 
and the small difference between eLLP and LDB is attributed to the smallness of the in-medium drag parameter $\eta$ in eLLP, 
and also to the evaluation of the induced fermion-fermion interactions 
(\ref{4f1}) and (\ref{4f2}). 
As for the interaction energy (the binding energy), 
it is approximated by the Hartree-Fock approximation in both methods, 
and in-medium effects appear as the Pauli blocking effects 
in the interaction integrals in Eq.~(\ref{recoil1}) and Eq.~(\ref{inte1}). 
The difference comes from the variational determination of 
the phonon momentum amplitude: 
the LDB leads to the recoil effect via $S(q)$ in Eq.~(\ref{fq1}), 
and the eLLP further includes the modification of kinetic energy of fermions 
through the $P$ dependence in $f_{q;P}$, 
and the drag effect by $\eta$ in addition to the overall blocking factor in Eq.~(\ref{fq2}). 

\begin{figure}[h] 
  \begin{center}
    \begin{tabular}{ccc}
\resizebox{80mm}{!}{\includegraphics{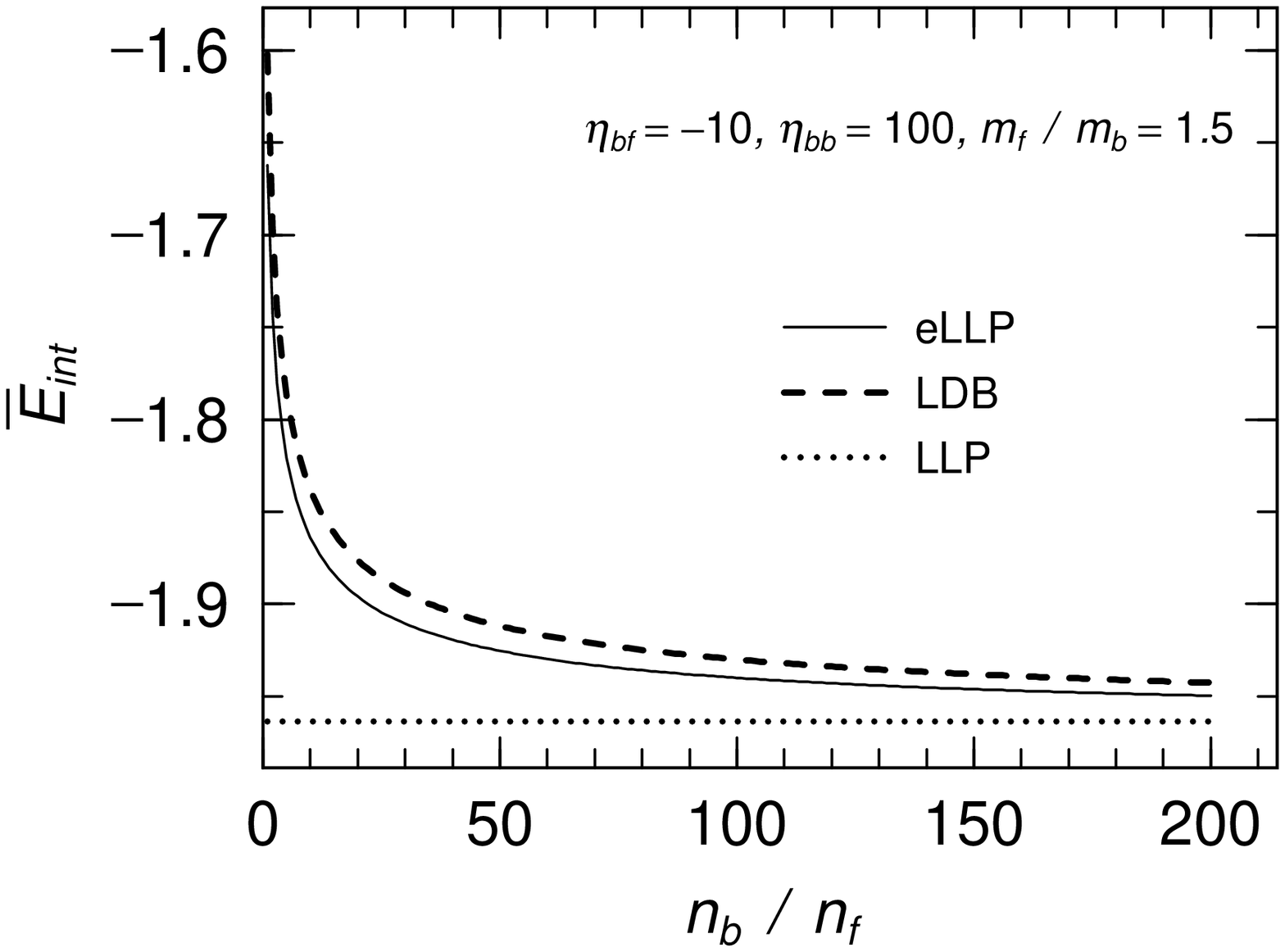}}
& \hspace{0.5cm} &
 \resizebox{76mm}{!}{\includegraphics{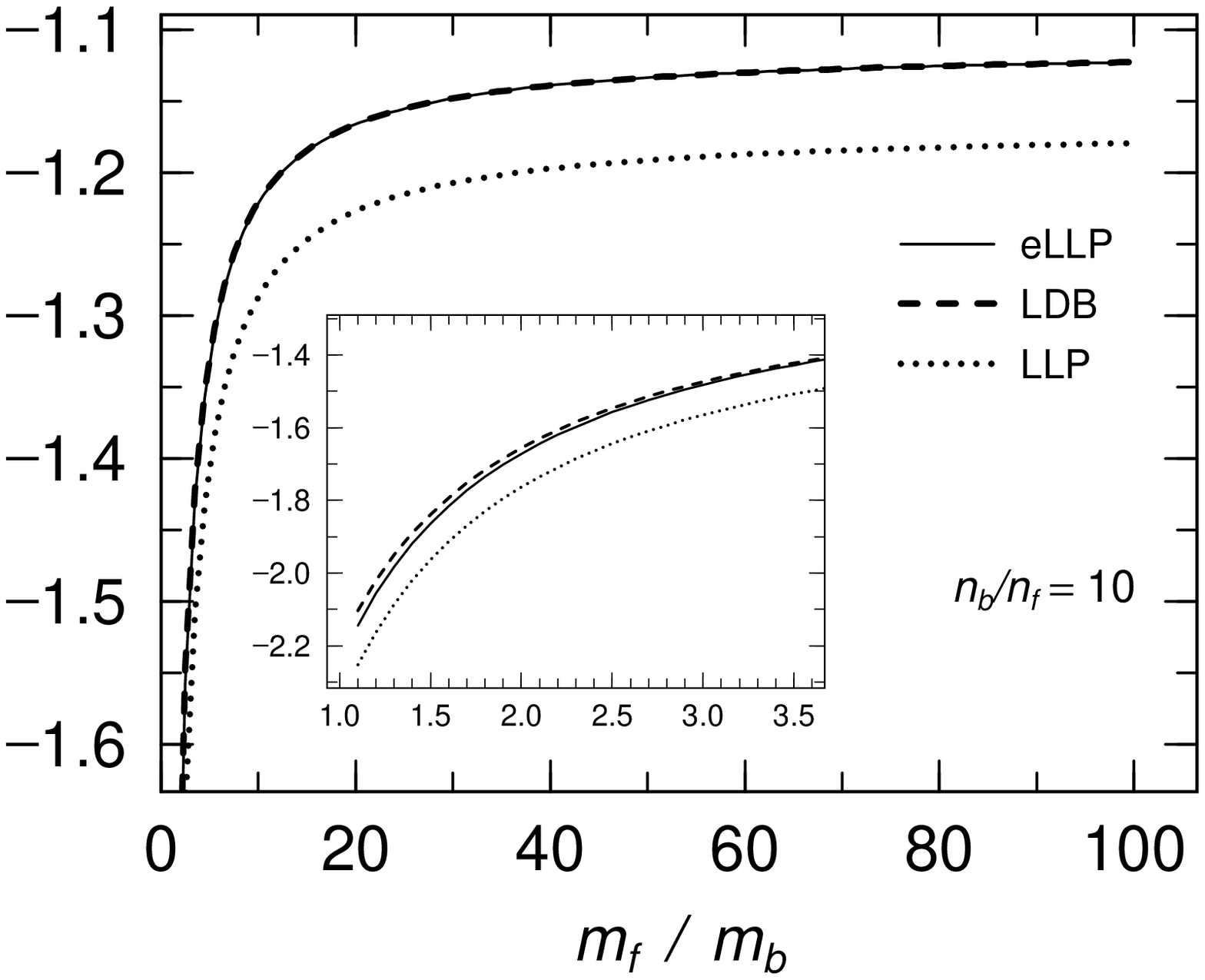}} \\
    \end{tabular}
    \caption{The interaction energies per fermion normalized as $\bar{E}_{int}=E_{int}/N_fE_0$ (\ref{EintenergyX}) 
    as functions of $Y=n_b/n_f$ (left) and $R=m_f/m_b$ (right). 
    The inset shows the blow-up of small $ R $ region. 
    Parameters are set as $\eta_{bf}=-10$, $\eta_{bb}=100$,  
    $R=1.5$ (left), and $Y=10$ (right). }
    \label{FigOur3}
  \end{center}
\end{figure}

Now we show 
the dependence of the ground state energy $E$ 
on the density and mass ratios, $n_b/n_f$ and $m_f/m_b$, both in eLLP and LDB.
Since the kinetic energy has the same form in these two methods, 
we evaluate the interaction energy per fermion in Eq.~(\ref{ELDB1}) and Eq.~(\ref{EOUR1}) defined by  
\beq
\frac{E_{int}}{N_f}=\frac{E}{N_f}-E_{kin}. 
\label{EintenergyX} 
\eeq
This quantity is comparable to the single polaron energy $E_{pol}(P)$ 
in (\ref{epol}) in LLP, 
and we set $P=0$ to have the interaction (binding) energy $E_{int}= E_{pol}(0)$ 
for the appropriate comparison in the dilute limit. 
Fig.~\ref{FigOur3} shows the scaled interaction energies  
as functions of the density and mass ratios, respectively;
the scaling of the energy is given in (\ref{scaling}).
As expected from the results presented above, 
the interaction energy approaches to the single-polaron LLP result in the dilute limit. 
On the other hand, 
the mass dependence (the right-panel of Fig.~\ref{FigOur3}) shows 
that the interaction energy converges to some asymptotic values in the heavy fermion limit: 
while the result in LLP approaches to the exact mean-field value as mentioned earlier, 
those in eLLP and LDB are different.
This is because the many-body effects still remain in the limit. 
These interaction energy (binding energy per fermion) can be measured 
from the radio-frequency absorption experiment \cite{Shashi1}, 
provided that the system is dilute and not so strongly-correlated 
for polarons to be identified as quasiparticles. 
%%%%%%%%%%%%%%%%%%%%%%%%%%%%%%%%%%%%%%%%%%%%%%%%%%%%%%%%%%%%%
%%%%%%%%%%%%%%%%%%%%%%%%%%%%%%%%%%%%%%%%%%%%%%%%%%%%%%%%%%%%%% 
%
\subsection{Drag parameter and effective mass}
%
%%%%%%%%%%%%%%%%%%%%%%%%%%%%%%%%%%%%%%%%%%%%%%%%%%%%%%%%%%%%%
\begin{figure}[h]
  \begin{center}
    \begin{tabular}{ccc}
 \resizebox{80mm}{!}{\includegraphics{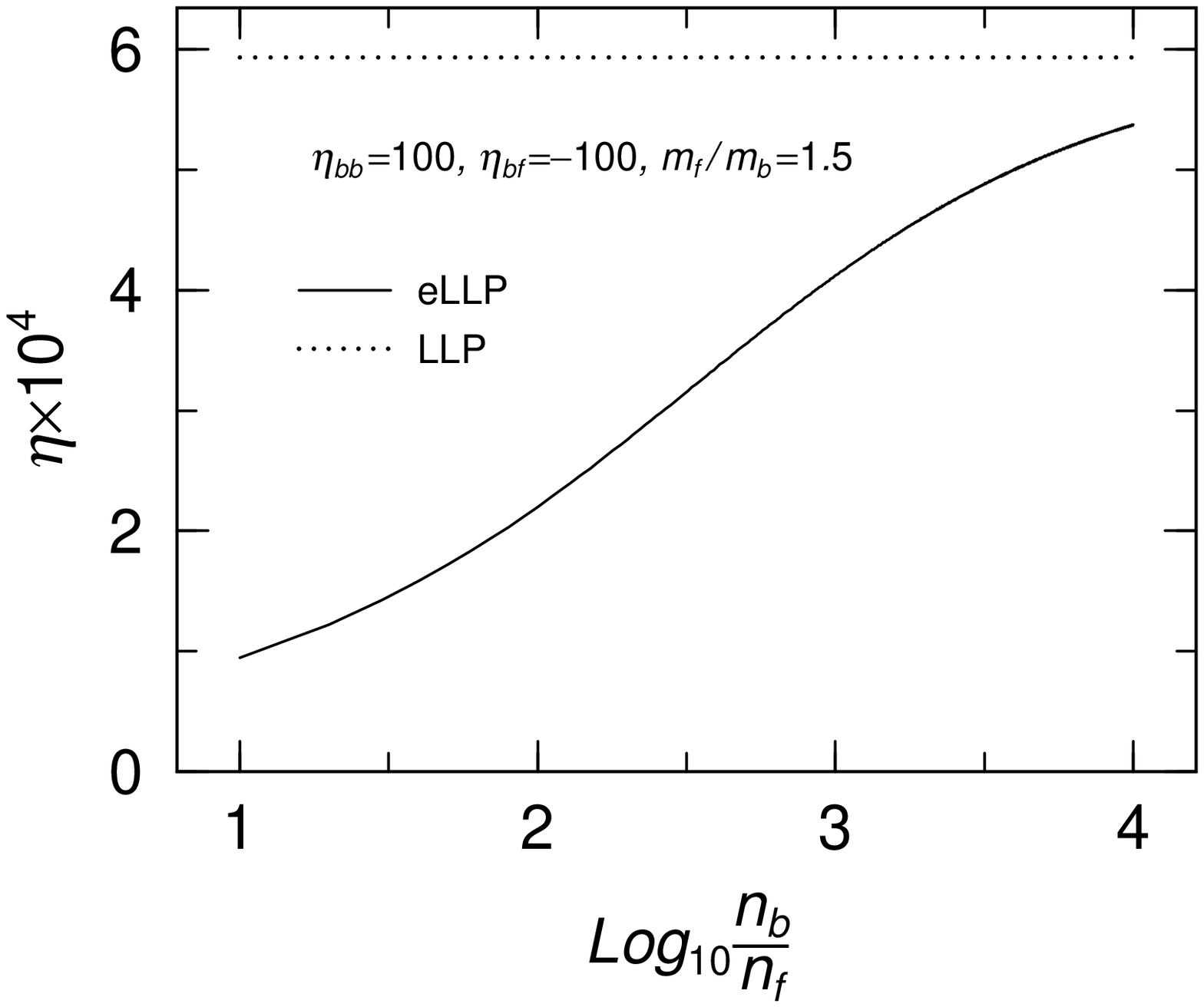}} 
 &\hspace{0.5cm}&
 \resizebox{90mm}{!}{\includegraphics{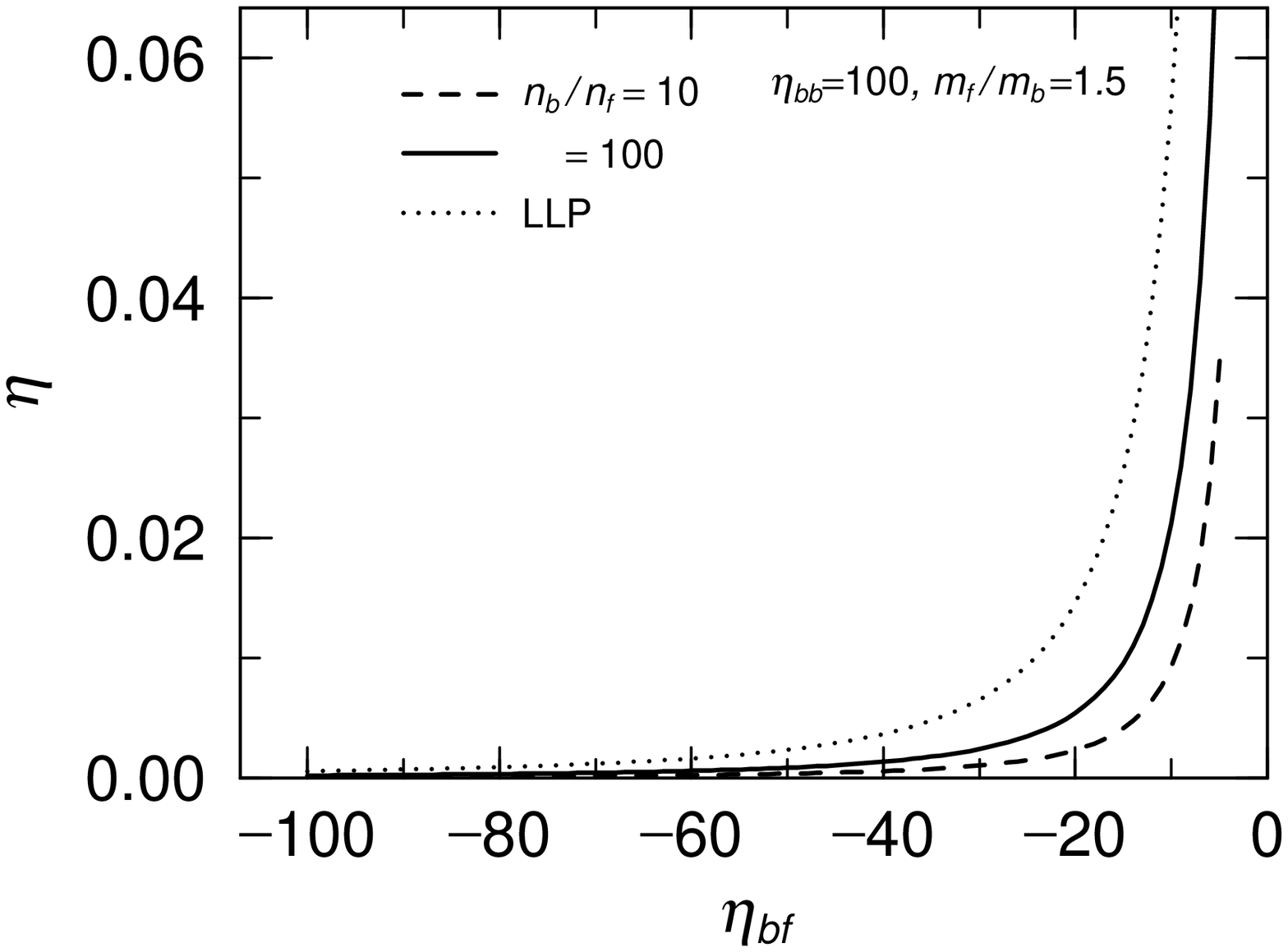}} 
    \end{tabular}
    \caption{Dependences of the drag parameter $\eta$ (\ref{eta1})
    on density ratio %$\log_{10}\frac{n_b}{n_f}$ 
    (left) and on boson-fermion coupling (right) % for $Y=n_b/n_f =10, 100, \infty$(LLP) 
    for $\eta_{bb} =100$ and $m_f/m_b =1.5$.}
    \label{FigOur2a}
  \end{center}
\end{figure}
%%%%%%%%%%%%%%%%%%%%%%%%%%%%%%%%%%%%%%%%%%%%%%%%%%%%%%%%%%%%%%  
As shown in Fig.~\ref{FigOur2a}, 
the drag parameter $\eta$ in the medium of fermions is very small 
and even smaller than that in the single-polaron LLP. 
It is due to the Pauli blocking effect of $f_{q;P}$ in Eq.~(\ref{imeta}).
Also, Fig.~\ref{FigOur2a} (left) shows that  
the in-medium effect weakens as the system is diluting 
and the result approaches to that in LLP in the dilute limit, and  
Fig.~\ref{FigOur2a} (right) that the increase rate of $\eta$ is very slow 
for the dimensionless boson-fermion scattering length inverse $\eta_{bf}$. 

%%%%%%%%%%%%%%%%%%%%%%%%%%%%%%%%%%%%%%%%%%%%%%%%%%%%%%%%%%%%%
\begin{figure}[h]
  \begin{center}
    \begin{tabular}{ccc}
\resizebox{75mm}{!}{\includegraphics{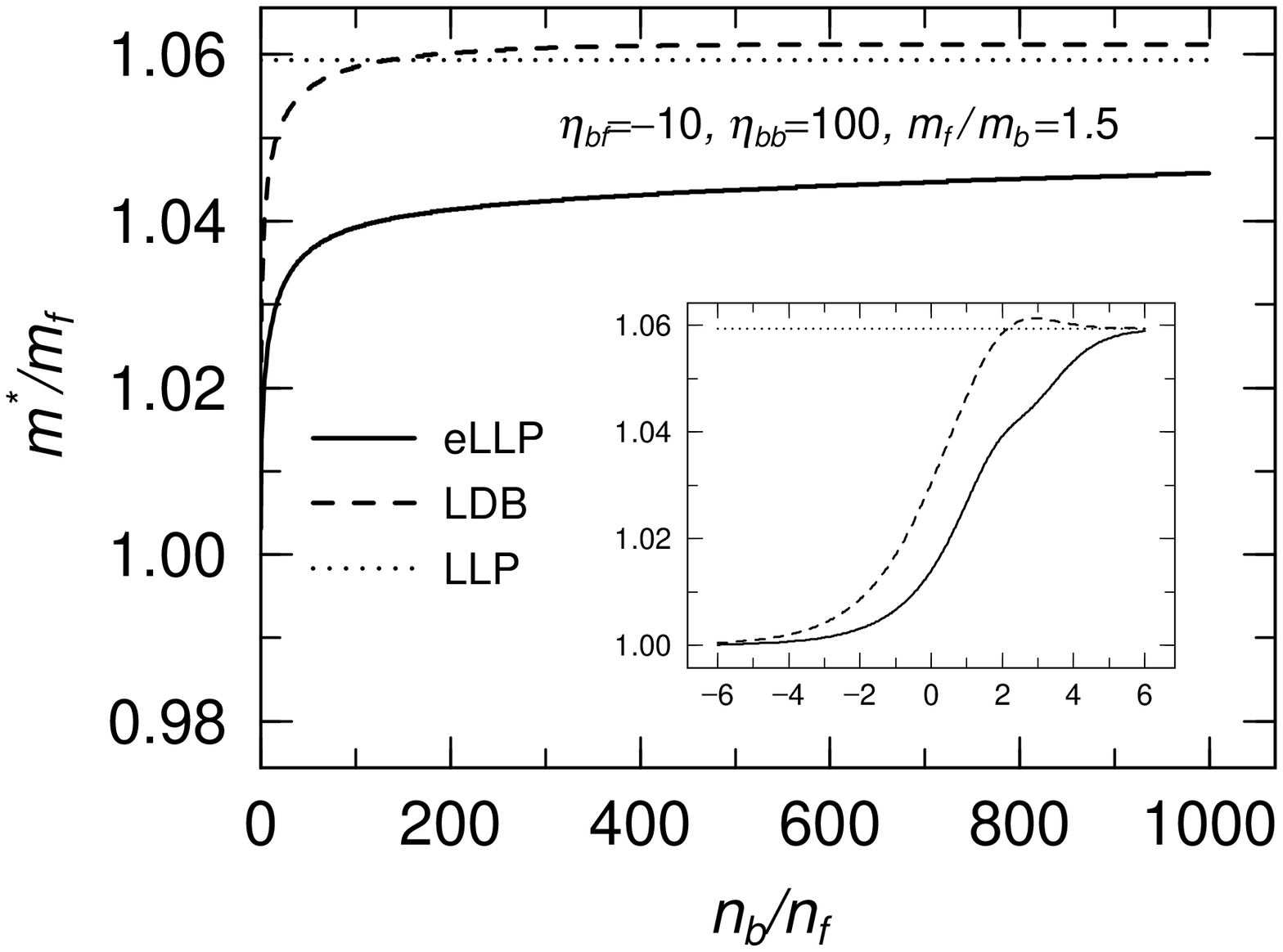}}
&\hspace{0.5cm}&
 \resizebox{78mm}{!}{\includegraphics{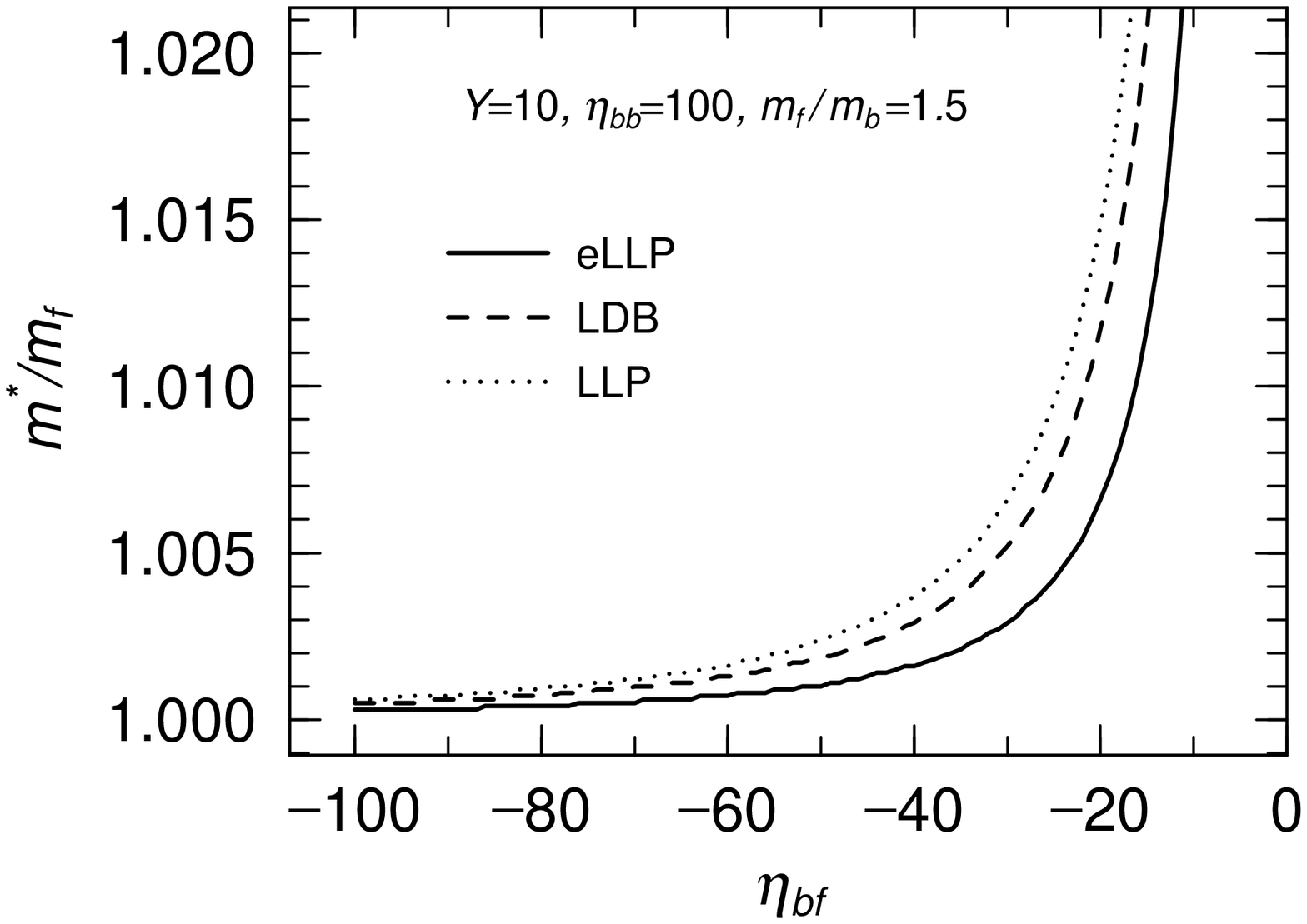}} 
    \end{tabular}
    \caption{Effective masses, (\ref{polaroneffmassX}), (\ref{eLLPmass}), and (\ref{LDBmass}), 
    as functions of the density ratio (left) for $\eta_{bf}=-10$ 
    and of the boson-fermion coupling (right) 
     for $Y=n_b/n_f =10$. 
     The inset figure (left) shows $m^*/m_f$ 
     as a function of $\log_{10}\frac{n_b}{n_f}$. 
    The other parameters are set by $R=m_f/m_b =1.5$ and $\eta_{bb}=100$. }
    \label{FigOur2}
  \end{center}
\end{figure}
%%%%%%%%%%%%%%%%%%%%%%%%%%%%%%%%%%%%%%%%%%%%%%%%%%%%%%%%%%%%%%
Fig.~\ref{FigOur2} shows the effective (inertial) masses in eLLP (\ref{eLLPmass}) 
and in LDB (\ref{LDBmass}) as functions of the density ratio 
and the boson-fermion coupling constant. 
It is noticed that the inertial and effective masses are different quantities: 
the former is a response function as the whole system to the external velocity, 
and the latter a curvature of polaronic dispersion relation.  
The mass in LDB is always larger than that in eLLP, 
and approaches to the LLP result from above (eLLP from below) in the dilute limit 
as shown in Fig.~\ref{FigOur2}(left).  
Although  in both cases the effective masses increase only by a few percent 
even for relatively large boson-fermion scattering lengths as shown in Fig.~\ref{FigOur2}(right), 
the many-body effects in the polaron gas seem to be significant 
when confronted with the case of the single polaron. 
As pointed out in \cite{LDB2,LDB4,LDB5}, 
these modifications in polaron masses in various situations can be measured 
from the Bragg spectroscopy 
for absorption and emission of the pair of laser beams by polarons. 

%%%%%%%%%%%%%%%%%%%%%%%%%%%%
\subsection{Comparison with perturbation theory}
%%%%%%%%%%%%%%%%%%%%%%%%%%%% 
The dilute gas of boson-fermion mixture being the same as the present case
has been studied in the 2nd order perturbation theory with respect to the boson-fermion coupling $g_{bf}$ 
in \cite{Viverit2}, where the Bogoliubov approximation is made, 
and the ground state energy is calculated in the form of 
$E^{pert}=N_fE_{kin}+E^{pert}_{int}$. 
The kinetic contribution $N_fE_{kin}$ is in common with ours, 
and the interaction one $E^{pert}_{int}$, 
which corresponds to the 2nd order sunset diagram of the fermion selfenergy with the Yukawa interaction, 
is given by 
\beq
\frac{E^{pert}_{int}}{N_fE_0}
&=&
\frac{2\pi a_{bf}}{m_{bf}}n_0 +\lk \frac{2\pi
a_{bf}}{m_{bf}}\rk^2 n_0 \int_q\frac{1}{q^2/2 m_{bf}}
\nn 
&-&
\lk \frac{2\pi a_{bf}}{m_{bf}} \rk^2 \frac{n_0}{n_f} 
\int_{q,P}\theta\lk q_F-|P|\rk 
\theta\lk |P-q|-q_F\rk 
\frac{\lk u_q-v_q\rk^2} 
{E_q+\frac{q^2}{2 m_f}
-\frac{q\cdot P}{m_f}},  
\label{Pert1}
\eeq
where we have rescaled variables in the equation (15) in \cite{Viverit2}, 
$A(\omega,\alpha)$, for comparison. 
The above equation (\ref{Pert1}) coincides with the eLLP result (\ref{EOUR1}) 
if the drag parameter $\eta$ is taken to be zero therein. 
In the previous section 
the drag parameter for many-body polarons was evaluated in the form of $\eta=(X+W)/(1+W)$ (\ref{eta1}); 
this result is non-perturbative 
since $X$ and $W$ are each of the order of $a_{bf}^2$ for the other parameters fixed.  
At the low density limit, the LLP theory is reproduced: $X\rightarrow 0$ and $W\rightarrow s$, 
where $s$ scales completely by $a_{bf}^2\lk n_0/a_{bb}\rk^{1/2}$ (\ref{ExFs}). 
Therefore, in the perturbative regime of the boson-fermion coupling,  
$\eta \sim {\mathcal O}\lk a_{bf}^2\rk$. 
The ground state energy (\ref{EOUR1}) in the eLLP method includes the $\eta$ in the denominator of the interaction integral 
and also in the modification of the kinetic energy $\sim \eta^2$, 
thus the eLLP method provides non-perturbative results via the mean-field type treatment of the phonon distribution function. 
Although the direct comparison between perturbative 
and non-perturbative results is not appropriate, 
the above observation shows that the eLLP method reduces 
to the 2nd order perturbation theory asymptotically 
in the small boson-fermion coupling limit. 
It is also interesting to note that the same is true on the LLP theory for the single polaron \cite{Grusdt2}. 

On the other hand, the LDB result (\ref{ELDB1}) seems to be of the order of $a_{bf}^2$, however, 
a many-body correlation effect comes in non-perturbatively through the structure factor $S(q)$. 
Actually, the LDB result reduces to the perturbative result in the low density limit. 
%%%%%%%%%%%%%%%%%%%%%%%%%%%%%
\subsection{Criterion for mean-field regime of polaron gas}
%%%%%%%%%%%%%%%%%%%%%%%%%%%%% 
As discussed in the literature \cite{Shashi1}, 
we can observe the breakdown of the mean-field regime 
in the momentum amplitude $f_{q;P}$ in LLP. 
Using (\ref{fqP}), 
the self-consistency condition 
$\eta P = \sum_{q \neq 0} q |f_{q;P}|^2$ becomes 
\beq
\eta &=&g_{bf}^2 n_0 \int_q \frac{q\cdot P}{|P|^2}
\frac{\varepsilon_q}{E_q}
\ldk E_q +\frac{q^2-2(1-\eta) q\cdot P}{2 m_f}\rdk^{-2}.
\eeq
This condition spoils 
when a singularity arises in the integral;
for the small $q$ region, 
the denominator is expanded, and, up to the linear order, 
it becomes 
\beq
 E_q +\frac{q^2-2(1-\eta) q\cdot P}{2 m_f} \simeq 
 |q|\lk v_{ph}-\frac{1-\eta}{m_f}|P|\cos\theta \rk, 
\eeq 
where $v_{ph}=\sqrt{4\pi a_{bb}n_0}/m_b$ is the sound velocity of phonon, 
and $\theta$ is an angle between $q$ and $P$.
Then, non-singularity condition gives the limitation for the
polaron velocity,
beyond which the mean-field solution does not work:
\beq
\frac{(1-\eta) |P|}{m_f} =\frac{|P|}{m_{eff}} < v_{ph}. 
\eeq
This is actually the signal that phonons are excited spontaneously 
by the interaction with the polaron faster than $v_{ph}$. 

For the polaron gas, we replace the momentum $P$ in the above condition 
with the Fermi momentum $p_F$, and obtain 
a similar criterion in terms of the boson-fermion density, 
mass ratios $n_0/n_f$, and $R=m_f/m_b$:
\beq
\frac{R}{1-\eta} \left( \frac{n_0}{n_f} \right)^{1/3}  >
\frac{(6 \pi^2)^{1/3}}{2\sqrt{\pi a_{bb}n_0^{1/3}}}
\sim 1.1\, \eta_{bb}^{1/2}. 
\eeq
Since the drag parameter $\eta$ varies from unity to zero with increasing $R$,
the mean-field-like approximation employed in this paper works 
for dilute and heavy mass regimes, 
which is consistent with the argument presented with the numerical results.

%%%%%%%%%%%%%%%%%%%%%%%%%%%%%%%%%%%%%%%%%%%%%
\section{Summary and outlook}
%%%%%%%%%%%%%%%%%%%%%%%%%%%%%%%%%%%%%%%%%%%%% 

We have studied the ground state properties of a
boson-fermion gaseous mixture at zero temperature,
where fermions are treated as a dilute gas of polarons in the BEC 
in the eLLP and the LDB methods,  
in which the unitary transformations are made for the eigenvalue problem 
of the system:
\beq 
H|\Psi\rangle &=& E |\Psi\rangle 
\ \rightarrow \ 
U^{-1}HU |\Psi'\rangle =  E |\Psi'\rangle, 
\eeq
where the ground state for $|\Psi'\rangle$ is approximated by 
the product of the phonon vacuum state 
and the Hartree-Fock ground state for fermions. 

It is found in both methods that the interaction energy per fermion,  
which should correspond to the binding energy of the single polaron in LLP, 
is suppressed by the many-body effects as the density of fermion is increased,  
but indeed becomes negative in relevant situations, i.e., 
for dilute and heavy fermions in the relatively weak coupling regime of the boson-fermion attraction. 
Also, we have found that the drag effect in eLLP is very small  
due to the many-body effect, and the difference of the ground state energy between eLLP and LDB
is not significant in the present approximations. 

For further studies, 
it is important to analyze 
the generalized unitary transformation (\ref{eLLPUni}) in more detail. 
In fact, this transformation generates higher order interaction terms among fermions, 
which we have truncated as they are expected to be negligible in the dilute gas of fermions, 
and kept only four-fermion interactions in the present study. 
This approximation seems to be valid 
because the higher-order interactions vanish 
when the $P$ dependence of the phonon momentum amplitude $f_{q;P}$ is negligible; 
the momentum $P$ is assigned to that of fermions 
and is of the order of the Fermi momentum at most. 
Since the $f_{q;P}$ is determined variationally for a given state, 
it is interesting to figure out in which region of the full parameter space  
the higher order interactions are controllable. 
Also, it is interesting to see how 
the perturbative corrections to the ground state modify results, 
since the transformed Hamiltonian $ U^{-1}HU $ includes different type interactions 
in eLLP and LDB. 

Also, methods presented here are applicable only to uniform and infinite systems. 
Further extensions to finite systems with discrete quantum states, 
such as in the harmonic trap potential or the optical lattice \cite{Compagno1,Hohmann1}, 
can be possible, 
as well as to possible inhomogeneity of the background BEC induced by polarons. 

The other interesting directions include 
study of the strong coupling regime using non-perturbative treatments. 
For this purpose we have to turn on the residual interactions  
having been dropped in the Bogoliubov approximation so far, 
which includes the boson-boson and boson-fermion interactions 
without condensation parts. 
These residual interactions account for 
many-body correlation effects beyond the present approximation, 
for instance, 
boson-fermion pair fluctuations develop to form composite fermion molecules 
around the unitarity limit of the boson-fermion attractive interaction. 
So it is interesting to observe such strong coupling effects 
on condensation fraction, 
modification of the polaron gas picture, 
spectral properties, and so on. 

%acknowledgement
%%%%%%%%%%%%%%%%%%%%%%%%%%%%%%%%%%%%%%%%%%%%% 
%%%%%%%%%%%%%%%%%%%%%%%%%%%%%%%%%%%%%%%%%%%%%

%%%%%%%%%%%%%%%%%%%%%%%%%%%%%%%%%%%%%%%%%%%%%%%%%%%%%%%%%%%
%\bibliography{bf_bec_refs-v1}
%%%%%%%%%%%%%%%%%%%%%%%%%%%%%%%%%%%%%%%%%%%%%%%%%%%%%%%%%%%

\appendix
%
%%%%%%%%%%%%%%%%%%%%%%%%%%%%%%%%%%%%%%%%%%%%%%%%%%%%%%%%%%%
\section{Bogoliubov approximation}\label{ApBog}
In this appendix, 
we present the Bogoliubov approximation for the effective Hamiltonian (\ref{Bogol1})
and derive the the Fr\"{o}hlich-type Hamiltonian (\ref{FrH}). 
The effective Hamiltonian (\ref{Hamil1}) provides the boson sector in the momentum representation:
\beq
H_b 
&=&
\sum_p \varepsilon_q b_q^\dagger b_q 
+
\frac{1}{V} \frac{1}{2} g_{bb} 
\sum_{k,p,q}
b_{p+q}^\dagger b^\dagger_{k-q} b_{k} b_{p},
\nonumber
\eeq
where the boson annihilation/creation operators, $b_p$ and $b^\dag_p$, 
are defined by $\phi(r) =V^{-1/2} \sum_p e^{ipr} b_p$.
Keeping the terms including the zero momentum component 
we obtain
\beq
H_{b;0} 
=
\sum_{q \neq 0} \varepsilon_q b_p^\dagger b_p 
+\frac{1}{2} g_{bb}\frac{N_0^2}{V}
+
\frac{1}{V} \frac{1}{2} g_{bb} 
\sum_{q\neq 0} 
\lk 
4b_{0}^\dagger b_{0} b^\dagger_{q} b_{q} 
+b_{0}^\dagger b_{0}^\dagger  b_{q} b_{-q} 
+b_{-q}^\dagger b_{q}^\dagger  b_{0} b_{0} 
\rk.
\nonumber
\eeq
In the case of the weak interaction at $T=0$,
the ground-state should be the BEC in the zero-momentum state of the condensed-particle number $N_0$, 
and we can use the approximation: $b_0, b_0^\dag \sim \sqrt{N_0}$. 
Then, the $H_{b;0}$ becomes
\beq
H_{b;0}
&\simeq& 
\frac{1}{2} g_{bb}\frac{N_b^2}{V}
+
\frac{1}{2} \sum_{q\neq 0} 
\lk \frac{q^2}{2 m_b}+n_0 g_{bb}\rk 
\lk  b_q^\dagger b_q+ b_{-q}^\dagger b_{-q} \rk
+
\frac{1}{2} n_0 g_{bb} 
\sum_{q\neq 0} 
\lk   
b_{-q}^\dagger b_{q}^\dagger
+b_{q}b_{-q}  
\rk 
%%%%%%%%%%%%%%%%%%%%%
\nn
&\simeq& 
\frac{1}{2} g_{bb}\frac{N_b^2}{V}+
\frac{1}{2}
\sum_{q\neq 0}
\lk
\begin{array}{cc}
b_{-q}^\dagger & b_{q} 
\end{array}
\rk 
\lk 
\begin{array}{cc}
\bar{\varepsilon}_q   &  g_{bb}n_0 \\
g_{bb}n_0         &  \bar{\varepsilon}_q 
\end{array}
\rk 
\lk
\begin{array}{c}
b_{-q} \\
b_q^\dagger 
\end{array} 
\rk 
-\frac{1}{2}\sum_{q \neq 0} \bar{\epsilon}_q,
\label{Hb02}
\eeq
where $N_b=N_0+\sum_{q\neq 0} b_q^\dagger b_q$ 
is the number operator of the boson;
$n_0=N_0/V$ the density of the condensed bosons, and  
$\bar{\varepsilon}_q=\varepsilon_q + g_{bb}n_0$. 

The matrix term in the last line of (\ref{Hb02}) 
is diagonalized with the quasi-particle annihilation/creation operators $C_q$, $C_q^\dagger$, 
which is defined by the Bogoliubov transformation:
\beq	
&&\lk
\begin{array}{c}
C_{-q} \\
C_q^\dagger 
\end{array} 
\rk 
=
\lk 
\begin{array}{cc}
u_q  &  v_q \\
v_q        &  u_q 
\end{array}
\rk
\lk
\begin{array}{c}
b_{-q} \\
b_q^\dagger 
\end{array} 
\rk, 
\label{BogTransf}
\eeq
where 
$u_q^2 =\frac{1}{2} \lk
1 +\frac{\bar{\varepsilon}_q}{E_q} \rk$
and  
$v_q^2=\frac{1}{2} \lk
-1+\frac{\bar{\varepsilon}_q}{E_q}\rk$
are the quasi-particle distribution functions, and 
$E_q=\sqrt{\varepsilon_q\lk \varepsilon_q + 2g_{bb}n_0 \rk }$ 
the quasiparticle energy of the Bogoliubov phonon. 
Then, we obtain the quasiparticle representation of $H_{b;0}$ in (\ref{Hb02}):
\beq
H_{b;0} \simeq
\frac{1}{2} g_{bb} \frac{N_b^2}{V}
+\frac{1}{2} \sum_{q \neq 0} \lk E_q- \bar{\varepsilon}_q \rk 
+\sum_{q \neq 0} E_q C_q^\dagger C_q,
\label{Hb0BT}
\eeq

Next, we evaluate the boson-fermion interaction term included in (\ref{Hamil1}): 
\beq
H_{int} 
&=&
\frac{1}{V} g_{bf} 
\int_r \psi^\dagger(r) \psi(r) 
\sum_{q,p} e^{i(p-q)r}b_{q}^\dagger b_p
%%%%%%%%%%%%%%%%%%%%%%%%%%%%%%%%%
\nn
&\simeq &
\frac{1}{V} g_{bf} N_0 \int_r \psi^\dagger(r) \psi(r)+
\frac{1}{V} g_{bf} N_0^{\frac{1}{2}}
\int_r \psi^\dagger(r) \psi(r) 
\sum_{q\neq 0} 
\lk e^{-i q \cdot r} b_{q}^\dagger + e^{i q \cdot r} b_q \rk
%%%%%%%%%%%%%%%%%%%%%%%%%%%%%%%%%
\eeq
where the boson-zero-momentum terms have been extracted.
Using the inverse Bogoliubov transformation for (\ref{BogTransf}):
\beq	
\lk
\begin{array}{c}
b_{-k} \\
b_k^\dagger 
\end{array} 
\rk 
&=&
\lk 
\begin{array}{cc}
u_k  &  -v_k \\
-v_k         &  u_k 
\end{array}
\rk 
\lk
\begin{array}{c}
C_{-k} \\
C_k^\dagger 
\end{array} 
\rk,
\eeq
we obtain the quasi-particle representation of the boson-fermion interaction term:
\beq
H_{int} 
&=&
\frac{1}{V} g_{bf} 
\int_r \psi^\dagger(r) \psi(r) 
\sum_{q,p} e^{i(p-q)r}b_{q}^\dagger b_p
%%%%%%%%%%%%%%%%%%%%%%%%%%%%%%%%%
\nn
&\simeq &
\frac{1}{V} g_{bf} N_0 \int_r \psi^\dagger(r) \psi(r)+
\frac{1}{V} g_{bf} N_0^{\frac{1}{2}}
\int_r \psi^\dagger(r) \psi(r) 
\sum_{q\neq 0} 
\lk e^{-i q \cdot r} b_{q}^\dagger + e^{i q \cdot r} b_q \rk
%%%%%%%%%%%%%%%%%%%%%%%%%%%%%%%%%
\nn
&=&
\frac{1}{V} g_{bf} N_0 N_f+
\int_r \psi^\dagger(r) \psi(r)  
\sum_{q\neq 0}
g_{q} \lk e^{-ir\cdot q} C_{q}^\dagger + e^{ir\cdot q} C_q\rk, 
\label{HintBogT}
\eeq
where the effective Yukawa coupling constant $g_q$ is defined by
\beq
g_q= \frac{1}{V} g_{bf} N_0^{\frac{1}{2}} \lk u_q-v_q\rk 
=\frac{1}{V} g_{bf} N_0^{\frac{1}{2}} \sqrt{\frac{\varepsilon_q}{E_q}}.
\label{YukawaBog}
\eeq 
Eqs. (\ref{Hb0BT}) and (\ref{HintBogT})  
provide the Fr\"{o}hlich-type Hamiltonian (\ref{FrH}) 
with the Yukawa coupling constant (\ref{YukawaC}) 
that corresponds to (\ref{YukawaBog}).

%%%%%%%%%%%%%%%%%%%%%%%%%%%%%%%%%%%%%%%%%%%
\section{Calculations on the parameter $s$}\label{ApCals}
%%%%%%%%%%%%%%%%%%%%%%%%%%%%%%%%%%%%%%%%%%%
%
In this appendix, 
we derive the explicit formula (\ref{ExFs}) for the parameter $s$.
Substituting (\ref{fqP}) into the self-consistency condition (\ref{Condeta}),
we obtain,
\beq
\eta P
&=& g_{bf}^2 n_0 \int_q q 
(u_q-v_q)^2 
\ldk E_q +\frac{q^2-2 q\cdot \lk P-P_{ph}\rk}{2 m_f}\rdk^{-2}, 
\eeq
Taking the leading-order term of the momentum $P$ in the right-hand-side integral,
we obtain the equation of $\eta$ for the small value of $P$: 
\beq 
\eta 
=
g_{bf}^2 n_0 \int_q 
(u_q-v_q)^2 \frac{2(1-\eta) \lk q\cdot P\rk^2}{m_f P^2}
\ldk E_q +\frac{q^2}{2 m_f}\rdk^{-3} +\mathcal{O}(P^3)
\eeq
In solving it by $\eta$, 
we represent the drag parameter $\eta$ as $\eta=s/(1+s)$, where  
\beq
s
&=&
\frac{2 g_{bf}^2 n_0}{m_f}
\frac{1}{(2\pi)^2}\int_0^\infty {\rm d}q q^4
\int_{-1}^1 {\rm d}x x^2
\frac{q^2}{2 m_bE_q} \ldk E_q +\frac{q^2}{2 m_f}\rdk^{-3}
\nn
&=&
\frac{32(1+R)^2}{3} \frac{a_{bf}^2 n_0^{\frac{1}{2}}}{a_{bb}^{\frac{1}{2}}}
\int_0^\infty {\rm d}z
\frac{z^2}
{\sqrt{z^2+16\pi}\lk
R\sqrt{z^2+16\pi}+z\rk^3}, 
\eeq
and we have used normalized variables in terms of 
$g_{bb}=4\pi a_{bb}/m_b$, $g_{bf}=2\pi a_{bf}/m_{bf}$
with the reduced mass $ m_{bf}=m_bm_f/\lk m_b+m_f \rk $, 
$ R=m_f/m_b $, and $z=q/\sqrt{a_{bb}n_0}$.  
%%%%%%%%%%%%%%%%%%%%%%%%%%%%%%%%%%%%%%%%%%%%%%%%%%%%%%%%%%%%%%%
\section{Interaction energy for two probe fermions}\label{PotproF}
%%%%%%%%%%%%%%%%%%%%%%%%%%%%%%%%%%%%%%%%%%%%%%%%%%%%%%%%%%%%%%%
We present the interaction energy between two heavy fermions placed at $r_1$ and $r_2$   
for the Fr\"{o}hlich-type Hamiltonian (\ref{FrH}). 
The interaction part of the Hamiltonian for the two probe fermions is described by 
\beq
H_1(r_1)+H_1(r_2)
&=&
\int_{q\neq 0}
g_q 
\ldk  e^{-ir_1\cdot q} C_{q}^\dagger 
+ e^{ir_1\cdot q} C_q\rdk 
+ (r_1 \rightarrow r_2),
\label{IntRR}
\eeq
and the second order perturbation theory 
gives the interaction energy between them: 
\beq
E^{(2)}(r_1,r_2)
&=&\sum_{q \neq 0}
2\frac{\langle 0| H_1(r_1)|q\rangle\langle
q|H_1(r_2)|0\rangle}
{E_0-E_q}, 
\label{SecPert}
\eeq
where $|0\rangle$ ($|q\rangle=C_q^\dagger|0\rangle$) 
is the Fock space for zero (single) phonon, 
and we have ignored the contribution of $E^{(2)}(r_{1(2)}, r_{1(2)})$ 
that corresponds to the second order selfenergy. 
Substituting (\ref{IntRR}) into (\ref{SecPert}),
we obtain 
\beq
E^{(2)}(r_1,r_2)
&=& 
-2 g_{bf}^2 \frac{N_0}{V^2} \sum_{q \neq 0} (u_q-v_q)^2 
\nn
&&\times 
\frac{\langle 0| \ldk  e^{-ir_1\cdot q} C_{q}^\dagger 
+ e^{ir_1\cdot q} C_q\rdk C_{q}^\dagger|0\rangle
\langle 0|C_{q} \ldk  e^{-ir_2\cdot q} C_{q}^\dagger 
+ e^{ir_2\cdot q} C_q\rdk|0\rangle}
{E_q}
%%%%%%%%%%%%%%%%%%%%%%%%%%%%%%
\nn
&=& 
-2 g_{bf}^2 n_0 \sum_{q \neq 0} \frac{(u_q-v_q)^2}{E_q}
e^{i(r_1-r_2)\cdot q}
\nn
&=& 
-\frac{g_{bf}^2 n_0 m_b}{i|r_1-r_2|\pi^2} 
\int_{-\infty}^\infty {\rm d}q \frac{q}{q^2+2\xi^{-2}}e^{i|r_1-r_2|q},
\nn 
&=& 
-\frac{g_{bf}^2 n_0 m_b}{\pi} \frac{e^{-|r_1-r_2|\sqrt{2}\xi^{-1}}}{|r_1-r_2|},
\label{IntRR2}
\eeq
where $\xi=1/\sqrt{8\pi n_0 a_{bb}}$ the coherence length.

%%%%%%%%%%%%%%%%%%%%%%%%%%%%%%%%%%%%%%%
\section{LDB transformation with anisotropic parameters}\label{ApLDBaI}
%%%%%%%%%%%%%%%%%%%%%%%%%%%%%%%%%%%%%%%
We present the result of the LDB transformation $U=e^{-S}$ 
with $S=-\int_r \hat{n}_f(r)Q(r)$, but, different from the original LDB,  
the momentum anisotropy exists in the momentum amplitude $f_q$:
\beq
f_q \neq  f_{-q}, \quad\hbox{and}\quad
\sum_{q \neq 0} q f_{q}\neq 0.
\eeq

\subsection{Transformation of field operators}
With $U =e^{-S}$, 
the fermion field operator $\psi(x)$ transforms
as
\beq
U^{-1} \psi(x) U 
=
e^{Q(x)} e^{W(x)} 
\psi(x), 
\qquad
U^{-1} \psi^\dagger(x) U
=
\psi^\dagger(x)
e^{W^*(x)}
e^{-Q(x)},
\label{psiTransf}
\eeq
where
\beq
W(x)\equiv \frac{1}{2}\ldk Q(x), -S\rdk 
%&=& \frac{1}{2}\ldk Q(x), -\int_r \hat{n}_f(r) Q(r)\rdk 
%
=
\frac{1}{2}\int_r \hat{n}_f(r) 
\int_{q} 
e^{i(x-r)\cdot q}\lk |f_q|^2 -|f_{-q}|^2 \rk, 
\label{Wx}
\eeq
We explain the derivation very shortly.
First, using the commutation relations,
\beq
\psi (-S)^n 
&=&
\ltk \ldk S,\psi \rdk - S \psi\rtk (-S)^{n-1}
=(Q-S)\psi S^{n-1} 
=(Q-S)^n\psi, 
%%%%%%%%%%%%%%%%%%%%%%%%
\\
0&=&\ldk \psi(x), Q(x')\rdk
=\ldk Q(x), \ldk S, Q(x')\rdk\rdk=\ldk S, \ldk S, Q(x)\rdk\rdk,
\eeq
the $U^{-1} \psi(x) U$ is transformed as
$U^{-1}\psi(x) U=
e^{S}\psi(x) e^{-S}
=e^{S} e^{Q(x)-S} \psi(x)$.
Then, using the Campbell-Baker-Hausdorff formula $e^X e^Y =e^Z$
with the operator $Z$:
\beq
Z&=&X+Y+\frac{1}{2}\ldk X,Y\rdk
+\frac{1}{12}\ldk X-Y,\ldk X,Y \rdk\rdk 
+\cdots, 
\nonumber
\eeq
for the $e^{S} e^{Q(x)-S}$, 
we obtain
%
%\beq
$U^{-1}\psi(x) U=
e^{Q(x)} e^{\frac{1}{2}\ldk S, Q(x)\rdk}\psi(x)$.
%\eeq
%
Direct calculation of the commutation relation $[Q(x),-S]$ proves Eqs. (\ref{psiTransf}) and (\ref{Wx}).

From Eq. (\ref{psiTransf}), 
we obtain the transformations of the derivatives of the fermion fields:
\beq
U^{-1}{\nabla}\psi(x)U
&=&{\nabla}(U^{-1}\psi(x)U)
\nn
&=& 
e^{Q(x)}e^{W(x)}  
\left\{ {\nabla} +{\nabla}Q(x)
-i \int_q q |f_q|^2
+{\nabla} W(x) \right\} 
\psi(x), 
%%%%%%%%%%%%%%%%%%%%%%%%%%%
\label{UDpsiU}
\\ 
U^{-1}{\nabla}\psi^\dagger(x)U
&=&{\nabla}(U^{-1}\psi^\dagger(x)U)
\nn
&=&
\psi^\dagger(x) \left\{ 
{\nabla}_L
-{\nabla} Q(x) 
+i \int_q q |f_q|^2
+{\nabla} W^*(x) \right\}
e^{W^*(x)}
e^{-Q(x)}, 
\label{UDpsiDU}
\eeq
where 
the $\nabla_L$ denote left derivative: $\psi^\dagger(x) \nabla_L =\nabla \psi^\dagger(x)$.
In the derivation of (\ref{UDpsiU}) and (\ref{UDpsiDU}), 
we have used the commutation relations:
\beq
\ldk \nabla W(x),W(x) \rdk =\ldk \nabla Q(x), W(x) \rdk =0, 
\eeq
and 
\beq
\nabla e^{Q(x)}
=
e^{Q(x)} \lk \nabla Q(x)-i\int_q q|f_q|^2 \rk. 
\eeq
The derivative formula of $e^{Q(x)}$ is proved 
from the next commutation relations:
\beq
\ldk \nabla Q(x),Q(x)\rdk
&=&-2 i \sum_{q \neq 0} q |f_q|^2, %\equiv -2iu, 
\\
\ldk \nabla Q(x),Q^n(x)\rdk
&=&
-i 2 n Q^{n-1}(x) \int_q q|f_q|^2, 
\eeq

The transformations of the phonon annihilation/creation operators 
are the same as the original LDB transformation:
\beq 
U^{-1}C_k U
&=&
C_k + f_k \int_r \hat{n}_f(r) e^{-ir\cdot k}, 
\\ 
U^{-1}C_k^\dagger U
&=&
C_k^\dagger + f_k^* \int_r \hat{n}_f(r) e^{ir\cdot k}. 
\eeq

\subsection{Transformation of Hamiltonian}
With the anisotropic LDB transformation,
the Fr\"{o}hlich-type Hamiltonian (\ref{FrH}) transforms as 
\beq 
U^{-1} \hat{H} U
&=&
U^{-1} \hat{H}_f U
+U^{-1} \hat{H}_{ph} U
+U^{-1} \hat{H}_{f{\rm\mathchar`-}ph} U,
\label{AiLDBham}
\eeq
where
\beq
U^{-1} \hat{H}_f U
&=&
\frac{1}{2 m_f} 
\int_x 
\left[\nabla \psi^\dagger(x)
-\psi^\dagger(x) 
\left\{ \nabla Q(x) -i \int_q q |f_q|^2 -\nabla W^*(x) \right\}
\right]
\nn 
&&\qquad\quad\times 
\left[ 
\nabla \psi(x)
+\left\{ \nabla Q(x) -i q \int_q |f_q|^2 +\nabla W(x) \right\} \psi(x) 
\right], 
%%%%%%%%%%%%%%%%%%%
\\
U^{-1} H_{ph} U 
&=&\int_{k} E_k 
\left\{ C_k^\dagger + f_k^* \int_r \hat{n}_f(r) e^{ir\cdot k} \right\}
\left\{ C_k + f_k \int_r \hat{n}_f(r) e^{-ir\cdot k} \right\}
%%%%%%%%%%%%%%%%%%%
\\
U^{-1}\hat{H}_{f{\rm\mathchar`-}ph}U
&=&
g_{bf} n_0^{\frac{1}{2}}
\int_r \hat{n}_f(r)
%\psi^\dagger(r) \psi(r) 
%
\int_{q}
( u_q -v_q ) 
\Big\{ e^{-ir\cdot q} C_q^\dagger %+ f_q^* \int_r \hat{n}_f(r) 
   +e^{ir\cdot q} C_q 
\nonumber\\
 & &\qquad\qquad\qquad\qquad\qquad\qquad
+( f_q^* + f_q ) \int_x \hat{n}_f(x) \Big\}
+g_{bf} n_0 \hat{N}_f. 
\eeq

Taking the normal ordering for the phonon operators,
we classify the terms of the Hamiltonian (\ref{AiLDBham}) by the order of fermion fields:
\beq
U^{-1} \hat{H} U
&=& H^{(2)}+H^{(4)}+H^{(6)}+H^{(no)}, 
\label{FNOH}
\eeq
where the second-order term $H^{(2)}$ is
\beq
H^{(2)}
&=&
\int_r \psi^\dagger(r) 
\Bigg\{
-\frac{1}{2 m_f}\nabla_r^2 
+\frac{1}{2 m_f}\lk \int_q q|f_q|^2 \rk^2 
+\int_q \lk E_q+\frac{q^2}{2 m_f}
+\frac{i q \cdot\nabla_r}{m_f}  \rk |f_q|^2 
\nn
& &\qquad\qquad
+g_{bf} n_0^{\frac{1}{2}}
\int_q \lk u_q-v_q\rk 
\lk 
f_{-q}^* + f_q \rk
\Bigg\}
\psi(r)
=
\int_P a_P^\dagger a_P E_{pol}(P),
\label{NOH2}
\eeq
where $E_{pol}(P)$ is defined as
\beq
E_{pol}(P) 
&=&
\frac{P^2}{2 m_f}
+
\frac{1}{2 m_f}\lk \int_q q|f_q|^2 \rk^2
+\int_q \lk E_q+\frac{q^2}{2 m_f}
-\frac{q\cdot P}{m_f} \rk |f_q|^2 
\nn
&+&
g_{bf} n_0^{\frac{1}{2}}
\int_q 
(u_q-v_q) 
(f_{-q}^* + f_q ).
\eeq
The forth-order and sixth-order terms are given by
\beq
H^{(4)}
&=&
-
\int_x \int_y 
\psi^\dagger(x)\psi^\dagger(y)\psi(x)\psi(y) 
\nn
& &\qquad\qquad\times 
\int_{q} 
e^{i(x-y)\cdot q}
\left\{ E_q|f_q|^2 +g_{bf} n_0^{\frac{1}{2}}
(u_q-v_q) (f_{-q}^* + f_q )
\right\} 
\nn
& &
-\frac{1}{2 m_f}
\left\{ 
\int_x 
\psi^\dagger(x) \nabla W(x) 
\cdot \lk \nabla -i\int_q q |f_q|^2 \rk \psi(x) + \hbox{h.c.}
\right\},
\\
H^{(6)}
&=&
\frac{1}{2 m_f}
\int_x 
\psi^\dagger(x) 
\left( \nabla W(x)^\dagger \cdot \nabla W(x) \right) \psi(x).
\eeq
The last one $H^{(no)}$ in (\ref{FNOH}) consists of the terms 
that include normal-ordered phonon fields 
such as $C_q^\dagger C_q$, $C_qC_q$, and $C_q^\dagger C_q^\dagger$. 

%%%%%%%%%%%%%%%%%%%%%%%%%%%%%%%%%%%%%%%
\section{Transformation of the extended LLP}\label{eLLPtra}
%%%%%%%%%%%%%%%%%%%%%%%%%%%%%%%%%%%%%%%
In this appendix, we present the transformations of boson and fermion field operators 
under the proposed unitary transformation $U=e^{S}$ given in (\ref{eLLPUni}).  
The fermion field transforms according to the similarity transformation formula 
\beq
U^{-1} \psi(x )U =\psi(x) 
+\ldk -S, \psi(x) \rdk 
+\frac{1}{2!}
\ldk -S, \ldk -S, \psi(x) \rdk \rdk +\cdots, 
\eeq
where the commutators are given by
\beq
\ldk -S, \psi(x) \rdk 
&=& %\int_y 
\alpha(x,y) \psi(y), 
\label{firstline1}
\\ 
\ldk -S,\ldk -S, \psi(x) \rdk \rdk 
&=& %\int_{y,z}\ldk 
w(x,y)\psi(y)+\alpha(x,y)\alpha(y,z)\psi(z), 
\\
-[S,[S,[S,\psi(x)]]] 
&=& -[S,w(x,y)]\psi(y)+2 w(x,y) \alpha(y,z) \psi(z) \nn
&&+ \alpha(x,y) w(y,z) \psi(z) 
+\alpha(x,y)\alpha(y,z)\alpha(z,u)\psi(u), 
\
\\
\vdots 
\nonumber
\eeq
where the abbreviation is used that the same arguments of space coordinates 
are integrated with, e.g., $\int_{y}$ in (\ref{firstline1}), 
and the operators $\alpha(x,y)$ and $w(x,y)$ defined in Eqs.~(\ref{alphaxy}-\ref{wxy}) 
include only phonon and fermion field operators, respectively.    
Now we approximate that $[S,w(x,y)]\simeq 0$ and $\alpha(x,y) w(y,z) \simeq w(x,y) \alpha(y,z)$ in the integral, 
since both of them generate higher order many-body interactions among fermions and phonons 
with anisotropic factors such as $|f_{k,P}|^2-|f_{k,-P}|^2$, 
which less contribute in the dilute limit of our interest. 
Then within these approximations the $n$-th commutation relation is given by 
\beq
[-S,\cdots,[-S,\psi(x)]\cdots]\equiv \langle x| \hat{X}_n |y \rangle \psi(y), 
\eeq 
where the operator $X_n$ in the bracket notation satisfies 
\beq
\hat{X}_n&=&\lk w \frac{\rm d}{{\rm d} \alpha} +\alpha \rk \hat{X}_{n-1} 
=\lk w \frac{\rm d}{{\rm d} \alpha} +\alpha \rk^n \hat{X}_0, 
\quad \mbox{with} \quad 
\hat{X}_{0}=1. 
\eeq
Using this, we obtain
\beq
U^{-1} \psi(x) U &=& \psi(x) +[-S,\psi(x)] + \frac{1}{2!}[-S,[-S,\psi(x)]]+ \cdots \nn
&& \simeq \sum_{n}^{\infty} \frac{1}{n!}\langle x| \hat{X}_n |y\rangle \psi(y)
\\
&=& 
\langle x| \exp{\lk w \frac{\rm d}{{\rm d} \alpha} +\alpha\rk} |y\rangle \psi(y)
=\langle x| \exp{\lk \alpha + \frac{1}{2}w \rk} |y\rangle \psi(y), 
\eeq
where the Baker-Campbell-Hausdorff formula $e^{A+B}=e^{A-\frac{1}{2}[A,B]}e^B$ has been used in the last line. 
Finally we obtain Eq.~(\ref{fertra1}) in the bracket notation. 

Next, using the transformations laws of boson and fermion field operators Eqs.~(\ref{fertra1}-\ref{phontra2}), 
we derive the transformation of the Hamiltonian Eq.~(\ref{FrHUnew}) as follows: 
The transformation of the fermion part $H_F'$ is given by 
\beq
H_F'&=& U^{-1} H_F U 
= 
\frac{1}{2m_f}\sum_P U^{-1}Pa_P^\dagger U \cdot U^{-1}P a_P U
\nn
%%%%%%%%%%%
&\simeq &
\frac{1}{2m_f}\sum_{P,Q,Q'}V^{-2} \int_{x,y,x',y'}
a_{Q}^\dagger e^{-iQy} \langle y| \hat{D}^\dagger e^{-\hat{A}} |x \rangle e^{iP(x-x')} 
\langle x'|  e^{\hat{A}} \hat{D} |y' \rangle e^{iQ'y} a_{Q'}
\nn
%%%%%%%%%%
&=&
\frac{1}{2m_f}\sum_{Q,Q'}V^{-1} \int_{y,y'}
a_{Q}^\dagger e^{-iQy} \langle y| \hat{D}^\dagger e^{-\hat{A}} e^{\hat{A}} \hat{D} |y' \rangle e^{iQ'y} a_{Q'}
\nn
%%%%%%%%%%
&=&
\frac{1}{2m_f}\sum_{Q,Q'}V^{-1} \int_{x,y,y'}
a_{Q}^\dagger e^{-iQy} D(x,y)^\dagger D(x,y) e^{iQ'y} a_{Q'}, 
\label{appHF}
\eeq
where the operator $\hat{D}$ and its representation are defined by 
\beq
\langle x| e^{\hat{A}} \hat{D} |y \rangle 
&\equiv&  -\nabla_x \langle x| e^{\hat{A}} |y \rangle 
\eeq
and 
\beq
D(x,y)&=& 
\langle x| \hat{D} |y \rangle =
V^{-1} \sum_{Q} e^{iQ(x-y)} \lk Q-u_Q-\sum_k e^{-ikx} X_{k,Q} \rk 
\\ 
\mbox{with } \quad 
u_Q &=& V^{-1}\sum_k k f_{k;Q}^2, \quad \mbox{and } \quad 
X_{k,Q}= V^{-1} \int_{x,y} e^{-i (Q-k) x} e^{i Q y} A(x,y).  
\eeq
Similarly the phonon part transforms as 
\beq
H_B'&=& U^{-1} H_B U = \sum_k E_k  U^{-1} C_k^\dagger U U^{-1} C_kU 
\nn 
&=&  \sum_k E_k \lk C_k^\dagger +\sum_P f_{k;P} a_P^\dagger a_{P-k} \rk  
 \lk C_k +\sum_P f_{k;P}^* a_{P-k}^\dagger a_{P} \rk, 
\label{appHB}
\eeq
and the interaction part as, in the momentum space, 
\beq
H_{I}'&=& U^{-1} H_{I} U 
= \sum_{k,P} g_k \lk U^{-1} C_k^\dagger  a_{P-k}^\dagger a_{P}  U + U^{-1} a_{P+k}^\dagger a_{P} C_k  U \rk 
\nn 
&\simeq & \sum_{k,P} g_k 
\lk C_k^\dagger +\sum_P f_{k;P} a_P^\dagger a_{P-k} \rk a_{P-k}^\dagger a_{P} 
+ h.c., 
\label{appHI}
\eeq 
where we have used the approximation $U^{-1} a_{P-k}^\dagger a_{P}U \simeq  a_{P-k}^\dagger a_{P}$ 
in the same manner for the fermion field operator. 
Finally, the sum of Eqs~(\ref{appHF}), (\ref{appHB}), and (\ref{appHI}) gives (\ref{FrHUnew}). 

%%%%%%%%%%%%%%%%%%%%%%%%%%%%%%%%%%%%%%%
\section{Transformation of Total Momentum Operator}\label{ApMom}
%%%%%%%%%%%%%%%%%%%%%%%%%%%%%%%%%%%%%%%
%
The total momentum operator $\hat{P}$ of the system consists of 
the phonon and the fermion parts:
\begin{equation*}
     \hat{P} =\hat{P}_F +\hat{P}_B
             =\sum_P P a^\dagger_P a_P +\sum_q q C^\dagger_q C_q.
\end{equation*}
The momentum is conserved 
in each terms in the unitary transformation (\ref{eLLPUni}), 
so that the momentum operator $\hat{P}$ is commutable with the transformation $U$: $[\hat{P},U] =0$,
from which we find that the total momentum operator $\hat{P}$ is invariant for the $U$-transformation:
\begin{equation}
     U^{-1} \hat{P} U =\hat{P}.
\label{Pinv}
\end{equation} 

Now, we calculate the $U$-transforms of the phonon and fermion parts, $\hat{P}_F$ and $\hat{P}_B$.
Using the transformation of $C_q$ and $C^\dagger_q$ in (\ref{phontra1}) and (\ref{phontra2}), 
we obtain
\beq
U^{-1} \hat{P}_B U &=& \sum_q q U^{-1} C^\dagger_q U U^{-1} C_q U \nn
                   &=& \sum_q q \left\{ C^\dagger_q +\sum_P f_{q;P} a_P^\dagger a_{P-k} \right\}
                                \left\{ C_q +\sum_P f_{q;P} a_{P-q}^\dagger a_P \right\}.
\nonumber
\eeq
Simple operator calculation gives the result:
\beq
U^{-1} \hat{P}_B U =\sum_q q C^\dagger_q C_q +\sum_{q,P} q \gamma_{q,P} a^\dagger_{P-q} a_P
                   +\sum_{q,P,Q} q f_{q;P} f_{q;P} b^\dagger_P a_{P-q} a^\dagger_{Q-k} a_Q,
\label{PBtrans}
\eeq
where 
\begin{equation*}
     \gamma_{q;Q} =f_{q;Q} C^\dagger -f_{-k;Q-k} C_{-k}.
\nonumber
\end{equation*}

Let's turn to the calculation of the fermion momentum operator $\hat{P}_F$.
Though direct calculation is difficult using the explicit form of the transformation $U$ in (\ref{eLLPUni}), 
we can obtain the exact result from the invariant relation (\ref{Pinv}):
\beq
     U^{-1} \hat{P}_F U &=& U^{-1} (P-\hat{P}_B) U 
                         = P -U^{-1} \hat{P}_B U.
\nonumber
\eeq
Using (\ref{PBtrans}), 
we obtain
\beq
     U^{-1} \hat{P}_F U 
     =\sum_P a^\dagger_P (P -u_P) a_P 
                -\sum_{q,P} q a^\dag_{P-k} \left\{ \gamma_{q;P} +\sum_Q \frac{d_{q,P,Q}}{2} a^\dagger_Q a_{Q-k} \right\} a_P,
\label{PFtran}
\eeq
where $u_P =\sum_k k f_{k;P}^2$ and 
$d_{q,P,Q} =f_{q;P} f_{q;Q} -f_{-q;P-q} f_{-q;Q-q}$.

It should be noted that the result (\ref{PFtran}) can also be obtained 
from the direct calculation of $U^{-1} \hat{P}_F U$ 
using the approximation (\ref{fertra1}).
It supports the use of (\ref{fertra1}) in the present calculations of eLLP.
%
%%%%%%%%%%%%%%%%%%%%%%%%%%%%%%%%%%%%%%%
\section{Normalization of interaction energies}
%%%%%%%%%%%%%%%%%%%%%%%%%%%%%%%%%%%%%%%
%%%%%%%%%%%%%%%%%%%%%%
\subsection{Binding energy in LLP}
%%%%%%%%%%%%%%%%%%%%%% 
We define the normalized form of the binding energy of the single polaron, 
which is defined by $E_{pol}(0)\equiv -E_{bin} (=E_{int})$ in Eq.~(\ref{epol}). 
The same renormalization procedure as the LDB method leads to 
the normalized binding energy: 
\begin{equation*}
-\bar{E}_{bin} =
\frac{4\pi (R+1)}{\eta_{bf}R} 
\Bigg[
1 
-\frac{2}{\pi \eta_{bf}} 
\int_0^\infty {\rm d}x 
%\lk
\left\{ 
\frac{x^2 (R+1)}
{R ( x^2+16\pi/\eta_{bb})
+\sqrt{x^2( x^2+16\pi/\eta_{bb})}}
-1\right\}
%\rdk
\Bigg], 
\end{equation*}
where $E_0$ is the boson zero-point energy. 

%%%%%%%%%%%%%%%%%%%%%%
\subsection{Interaction energy in LDB}
%%%%%%%%%%%%%%%%%%%%%%
The interaction energy is defined as 
$E_{int} =E -N_f E_{kin}$, 
where $E$ is the renormalized ground-state energy in (\ref{ELDB1}). 
The normalized energy $\bar{E}_{int}$, which is scaled with the boson zero-point energy $E_0$
and the fermion number $N_f$, 
is represented by 
\beq
&&\frac{{\bar{E}}_{int}}{N_f} \equiv
\frac{E_{int}}{E_0 N_f}
%\frac{E-N_f E_{kin}}{N_f E_{0}}
%
=\frac{4\pi (R+1)}{\eta_{bf} R} \nn
&&\times 
\Bigg[ 
1 
-\frac{2}{\pi \eta_{bf}} 
\int_0^\infty {\rm d}x 
\left\{ 
\frac{S^2(x) x^2 (R+1)}
{R S(x)\lk x^2+16\pi/\eta_{bb}\rk
+\sqrt{x^2\lk x^2+16\pi/\eta_{bb}\rk}}
-1\right\}
\Bigg], 
\nn 
\label{LDBapp}
\eeq
where $Y\equiv n_0/n_f$, $x\equiv q/n_0^{1/3}$, and 
\beq
S(x)
&=&
\frac{x}{2}\lk \frac{Y}{6\pi^2}\rk^{\frac{1}{3}}
\left\{
\frac{3}{2}
-\frac{x^2}{8} \lk
\frac{Y}{6\pi^2}\rk^{\frac{2}{3}}
\right\} 
\theta\left[\lk \frac{6\pi^2}{Y}\rk^{\frac{1}{3}}-\frac{x}{2}\right]%
%\nn
%&&
+
\theta\left[ \frac{x}{2}-\lk \frac{6\pi^2}{Y}\rk^{\frac{1}{3}}\right].
\eeq

%%%%%%%%%%%%%%%%%%%%%%
\subsection{Interaction energy in eLLP}
%%%%%%%%%%%%%%%%%%%%%%
%
The interaction energy in eLLP is defined in the same way as in LDB;
it is obtained from  (\ref{EOUR1}) as $E_{int}=E-E_{kin}N_f$.
The interaction energy $E_{int}$ consists of three parts:
$E_{int}=E_{mf}+E^{(1)}_{int}+E^{(2)}_{int}$, 
The $E^{(1)}_{int}$ and  $E^{(2)}_{int}$ are the contributions from 
$0 \le q \le 2q_F$ and $ q\ge 2q_F$, respectively:
\beq
\frac{E_{int}^{(1)}}{N_f}
&=&
-
\lk \frac{2\pi a_{bf}}{m_{bf}} \rk^2 \frac{n_0}{n_f} 
\int_{q,P}
\theta( 2 q_F -|q| ) 
\theta( q_F -|P| ) 
\theta( |P-q| -q_F ) \nn
&&\qquad\qquad\qquad\qquad\qquad\qquad\times
\frac{\lk u_q-v_q\rk^2} 
{E_q+\frac{q^2}{2 m_f}
-\frac{q\cdot P(1-\eta)}{m_f}}
%%%%%%%%%%%%%%%%%%%%
\nn
&=&
-\frac{\lk 1+R \rk^2 Y E_0}{2\pi^2 \eta_{bf}^2 R}
\int_0^{2 x_F} 
\frac{{\rm d}x }{1-\eta} \nn
&& \times \Bigg[
\lk C+x_F\rk 
-\frac{x}{2}
-\frac{1}{x}\lk C^2-x_F^2\rk 
\ln\lk\frac{C+x_F}{C+x_F-x}\rk 
\nn
&&
\qquad\qquad\qquad\qquad\qquad
+\lk x-2C\rk 
\ln\lk\frac{C+x_F-x}{C-x/2}\rk 
\Bigg]
\frac{x^3}{\sqrt{x^2+16\pi/\eta_{bb}}}, 
\eeq
and 
\beq 
\frac{E_{int}^{(2)}}{N_f}
&=&
-
\lk \frac{2\pi a_{bf}}{m_{bf}} \rk^2 \frac{n_0}{n_f} 
\int_{q,P}\theta\lk |q|-2 q_F\rk \theta\lk q_F-|P|\rk  
\frac{\lk u_q-v_q\rk^2} 
{E_q+\frac{q^2}{2 m_f}
-\frac{q\cdot P(1-\eta)}{m_f}}
%%%%%%%%%%%%%%%%%%%%
\nn
&=&
-
\lk \frac{2\pi a_{bf}}{m_{bf}} \rk^2 \frac{n_0^3}{n_f} 
\frac{8\pi^2}{(2\pi)^6}\int_{2 x_F}^{\infty}
\frac{ {\rm d}x x^3}{\sqrt{x^2+16\pi/\eta_{bb}}}
\int_0^{x_F} {\rm d}y y^2 \int_{-1}^{1}{\rm d}z 
\ldk 
\frac{1}{A-B P z}
\rdk 
%%%%%%%%%%%%%%%%%%%%
\nn
&=&
-\frac{\lk 1+R \rk^2 Y E_0}{\pi^2 \eta_{bf}^2 R}
\int_{2 x_F}^{\infty}
\frac{ {\rm d}x x^2/(1-\eta)}{\sqrt{x^2+16\pi/\eta_{bb}}}
\ldk C x_F+\frac{x_F^2-C^2}{2}\ln\left(\frac{C+x_F}{C-x_F}\right) \rdk, 
\eeq
%%%%%%%%%%%%%%%%%%%%%%%%%%%%%%%%%%%%%%%%%%%%%%%%%%%%%%%%%
%%%%%%%%%%%%%%%%%%%%%%%%%%%%%%%%%%%%%%%%%%%%%%%%%%%%%%%%%
%%%%%%%%%%%%%%%%%%%%%%%%%%%%%%%%%%%%%%%%%%%%%%%%%%%%%%%%%
where $x=q/n_0^{1/3}$ and $(u_q-v_q)^2=\frac{x}{\sqrt{x^2+16\pi/\eta_{bb}}}$.
In the integration of the momentum $P$,
we have used the cylindrical coordinates $(P_T,\phi,P_z)$ where $q \parallel P_z$. 
For $0\le q \le 2q_F$, 
the integration of the radial coordinate $P_T$ has been done first using the formula:
\beq
&&\int {\rm d}^3P \ 
\theta\lk q_F-|P|\rk \theta\lk |P-q|-q_F\rk \nn
&&\quad =
-2\pi \int_{q/2}^{-q_F+q}{\rm d}P_z
\int_{\sqrt{q_F^2-(q-P_z)^2}}^{\sqrt{q_F^2-P_z^2}}
{\rm d}P_T P_T 
-2\pi \int_{-q_F+q}^{-q_F}{\rm d}P_z
\int_0^{\sqrt{q_F^2-P_z^2}}
{\rm d}P_T P_T
\nn
&&\quad=
\pi \int_{q/2}^{-q_F}{\rm d}P_z
P_z^2
-
\pi \int_{q/2}^{-q_F+q}{\rm d}P_z
(q-P_z)^2
-
\pi q_F^2 \int_{q-q_F}^{-q_F}{\rm d}P_z, 
\eeq
Finally, the integral of the momentum $P$ for $0\le q \le 2q_F$ becomes 
\beq
&&\frac{1}{\pi} \int {\rm d}^3P \theta\lk q_F-|P|\rk \theta\lk |P-q|-q_F\rk
\frac{1}{A-B P_z}
\nn
&&\quad =
-
\left[ \frac{P_z^2}{2B}
+\frac{A}{B^2}\lk P_z+\frac{A}{B}\ln(A-BP_z)\rk \right]_{q/2}^{-q_F}
+ 
\frac{q_F^2}{B}
\left[ \ln(A-BP_z)
\right]_{q-q_F}^{-q_F}
\nn
&&\qquad+ 
\frac{1}{B}
\left[ \lk 2 q+\frac{A}{B}\rk P_z +\frac{P_z^2}{2}
+\lk q+\frac{A}{B}\rk^2\ln(A-BP_z)
\right]_{q/2}^{q-q_F}
%%%%%%%%%%%%%%%%%%%%%%%%%
\nn
&&\quad=
\frac{q}{B}\lk \frac{A}{B}+q_F\rk -\frac{q^2}{2B}
-\frac{1}{B}\lk \frac{A^2}{B^2}-q_F^2\rk 
\ln\lk\frac{A+Bq_F}{A+B(q_F-q)}\rk
\nn
&&\qquad
+\frac{q}{B}\lk q-\frac{2A}{B}\rk \ln\lk\frac{A+B(q_F-q)}{A-Bq/2}\rk
%%%%%%%%%%%%%%%%%%%%%%%%%
\nn
&&\quad=
\frac{m_f n_0^{1/3}}{1-\eta}
\Bigg\{ 
(C +x_F) 
-\frac{x}{2}
-\frac{1}{x} (C^2-x_F^2) 
\ln\lk\frac{C+x_F}{C+x_F-x}\rk
\nn
&&\qquad\qquad\qquad\qquad\qquad\qquad\qquad
+(x-2C) 
\ln\lk\frac{C+x_F-x}{C-x/2}\rk 
\Bigg\}, 
\eeq
where
\begin{equation*}
A=E_q+\frac{q^2}{2 m_f}, \quad 
B=\frac{q(1-\eta)}{m_f}, 
\quad
\frac{A}{B}=
\frac{R\sqrt{x^2+16\pi/\eta_{bb}}+x}{2(1-\eta)}n_0^{1/3}
\equiv C n_0^{1/3}.  
\end{equation*}
In the case of $q \geq 2q_F$,
we make the replacement  
$\theta\lk |P-q|-q_F\rk \rightarrow 1$ in the integral, 
and the integrand of the momentum $P$ integral becomes spherically symmetric. 

It should be noted that we apply the renormalization procedure to the divergence of $E_{int}^{(2)}$ 
in the same manner as in the LLP and the LDB.

%%%%%%%%%%%%%%%%%%%%%%
\end{document}